\documentclass[journal]{IEEEtran}
\ifCLASSINFOpdf
   \usepackage[pdftex]{graphicx}
\else
\fi
\usepackage{amssymb}
\usepackage{multirow}
\usepackage{subfigure}
\usepackage{float}
\usepackage{epstopdf}
\usepackage{amsmath}
\usepackage{array}
\begin{document}
\title{CCL: Cross-modal Correlation Learning with Multi-grained Fusion by Hierarchical Network}

\author{Yuxin Peng, Jinwei Qi, Xin Huang and Yuxin Yuan
\thanks{This work was supported by National Natural Science Foundation of China under Grants 61371128 and 61532005.}
\thanks{The authors are with the Institute of Computer Science and Technology,	Peking University, Beijing 100871, China. Corresponding author: Yuxin Peng	(e-mail: pengyuxin@pku.edu.cn).}
}

\markboth{IEEE TRANSACTIONS ON MULTIMEDIA}%
{Shell \MakeLowercase{\textit{et al.}}: Bare Demo of IEEEtran.cls for IEEE Journals}



\maketitle

\begin{abstract}
	%Cross-modal retrieval has become a highlighted research topic, which can provide flexible retrieval experience across multimedia data such as image, video, text and audio. 
	Cross-modal retrieval has become a highlighted research topic for retrieval across multimedia data such as image and text. 
	%Recently, researchers attempt to explore cross-modal retrieval via Deep Neural Network (DNN), and 
	A two-stage learning framework is widely adopted by most existing methods based on Deep Neural Network (DNN): \textit{The first learning stage} is to generate separate representation for each modality, and \textit{the second learning stage} is to get the cross-modal common representation. 
	However, the existing methods have three limitations: (1) In \textit{the first learning stage}, they only model intra-modality correlation, but ignore inter-modality correlation with rich complementary context. % for learning better separate representation. 
	(2) In \textit{the second learning stage}, they only adopt shallow networks with single-loss regularization, 
	but ignore the intrinsic relevance of intra-modality and inter-modality correlation. %, so cannot effectively exploit and balance them to improve generalization performance. 
	%and cannot effectively balance intra-modality and inter-modality correlation, where the relative importance between them is not stable and needs adaptive tuning; 
	(3) Only original instances are considered while the complementary fine-grained clues provided by their patches are ignored.
	For addressing the above problems, this paper proposes a cross-modal correlation learning (CCL) approach with multi-grained fusion by hierarchical network, and the contributions are as follows:
	(1) In \textit{the first learning stage}, CCL exploits multi-level association with joint optimization to preserve the complementary context from intra-modality and inter-modality correlation simultaneously. %, while existing methods generate separate representation only with intra-modality information. 
	(2) In \textit{the second learning stage}, a multi-task learning strategy is designed to adaptively balance the	intra-modality semantic category constraints and inter-modality pairwise similarity constraints. %, while existing methods generate common representation with shallow networks of single-loss regularization.
	(3) CCL adopts multi-grained modeling, which fuses the coarse-grained instances and fine-grained patches to make cross-modal correlation more precise. %, while the existing methods only focus the original instances.
	Comparing with 13 state-of-the-art methods on 6 widely-used cross-modal datasets, the experimental results show our CCL approach achieves the best performance.
	
\end{abstract}


% \fi
\IEEEpeerreviewmaketitle

\section{Introduction}

\begin{figure}[!t]
	\centering
	\includegraphics[width=0.4\textwidth]{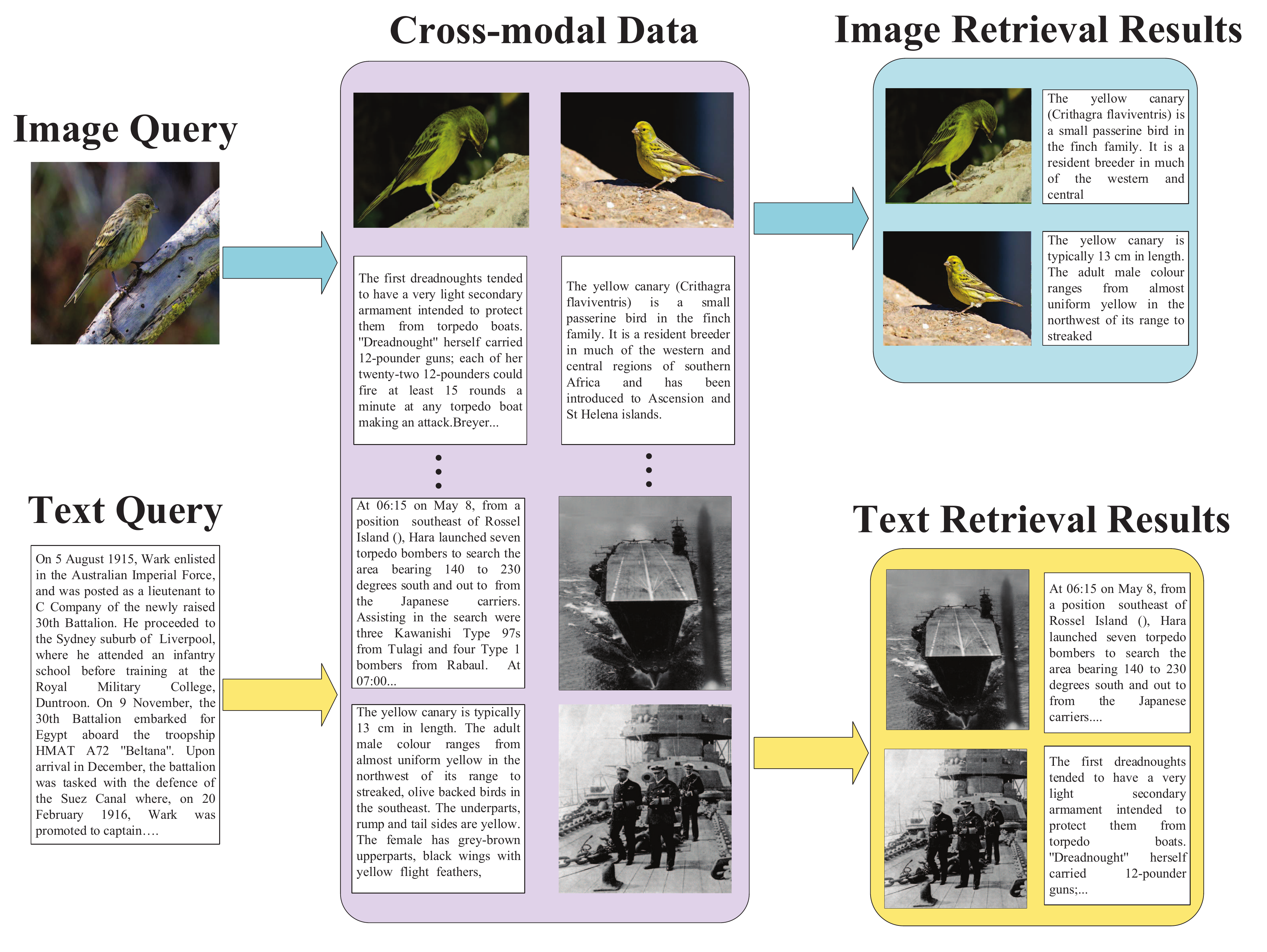}
	\caption{An example of cross-modal retrieval with image and text, which can present retrieval results with different modalities.}
	\label{fig_cross_media}
\end{figure}

With rapid development of computer science and technology, %information acquisition and processing have been transformed from the form of single media to multimedia, such as image, video, text and audio. 
multimedia data including image, video, text and audio, has been emerging on the Internet and reshaping people's life. Consequently, multimedia retrieval has been an essential technique with wide applications, such as search engine and multimedia data management.
The traditional retrieval methods mainly focus on single-modal scenario \cite{HuTMM2009Coherent, PengCSVT06ClipRetrieval}, which provides retrieval results of the same single modality with query, such as image retrieval and text retrieval. Furthermore, some methods attempt to address the retrieval problem where multimedia data exists as tight combination \cite{ZnaidiaICPR2012bag,LiuCIVR2010coherent}. 
But the major limitation of these methods is that the retrieval results must share the same modality combination with user's queries, e.g. retrieving image/text pairs with the query of an image/text pair. They cannot directly measure the similarity between different modalities, which restricts the retrieval flexibility. 
%For example, in the method of [], users submit a query of x-y pair to retrieve the relevant x-y pairs in the dataset. That is to say, the multimodal similarity is a direct combination of x-x and y-y similarity, but cross-modality similarity as the similarity between image and audio cannot be obtained. 
%In a word, The above methods have the limitation on the flexibility of modalities, which restricts the flexibility of information retrieval. 

Cross-modal retrieval is a relatively new retrieval paradigm, which can perform retrieval across multimedia data. For example, if someone is interested in canary, he can submit one image query, and then get relevant multimedia information, including text descriptions, image samples, video introductions, audio clips and so on. Figure \ref{fig_cross_media} is an example of cross-modal retrieval with image and text. Compared with single-modal retrieval, cross-modal retrieval can provide more flexible and useful retrieval experience to show rich multimedia search results.
The key problem of cross-modal retrieval is that the distribution and representation of different modalities are inconsistent, and such ``heterogeneity gap'' makes it hard to measure the cross-modal similarity. 

For bridging ``heterogeneity gap'', %``modality gap'', 
%some early methods \cite{DBLP:journals/tmm/ZhuangYW08,yang2008harmonizing} construct correlation graphs to perform cross-modal retrieval, where the cross-modal data exist as multimedia documents (MMDs). 
%However, they are still not flexible enough because in practical scenario, cross-modal data usually exist separately instead of as MMDs. 
most existing methods are proposed to learn a common space for different modalities. %, as shown in Figure \ref{fig_CommonSpace}. 
These methods like \cite{RasiwasiaMM10SemanticCCA, DBLP:journals/tmm/DarasMA12, zhang2016cross} aim to project the features from single-modal space into cross-modal common space and get common representation for similarity measure.
The common representation is usually generated following principles such as maximizing cross-modal pairwise correlation \cite{RasiwasiaMM10SemanticCCA}, maximizing classification accuracy in the common space \cite{ZhaiTCSVT2014JRL}, etc. %and preserving similarity constraints from the single-modal space, etc. 
According to the different adopted models, existing methods can be divided into two major ways. The first is to learn linear projections in traditional frameworks, like Canonical Correlation Analysis (CCA) \cite{RasiwasiaMM10SemanticCCA} and graph-based methods \cite{ZhaiTCSVT2014JRL}. 
However, their performance is limited by the traditional framework, which cannot capture the complex cross-modal correlation with high non-linearity. As indicated in \cite{DBLP:conf/icml/AndrewABL13}, although some kernel based methods have the ability of learning non-linear representation, the learned representation is limited due to the fixed kernel. 
With the great progress by deep learning in single-modal scenario such as image classification, there arises the second way for common representation learning, which takes DNN as the basic model. These methods take the advantage of DNN's strong ability of non-linear modeling for analyzing complex cross-modal correlation \cite{feng12014cross,DBLP:conf/ijcai/PengHQ16,DBLP:conf/cvpr/YanM15}, which avoid the aforementioned problems of nonparametric models in traditional methods to learn more flexible non-linear representation.% as indicated in \cite{DBLP:conf/icml/AndrewABL13}.

DNN-based methods for common representation learning can be mainly divided into two stages: \textit{The first learning stage} is to generate separate representation for each modality, and \textit{the second learning stage} is to learn common representation by exploiting cross-modal correlation.
However, in the first learning stage, these methods only model intra-modality correlation to obtain separate representation \cite{srivastava2012learning}, but ignore the intrinsic correlation within inter-modality. 
%, which can provide rich complementary context to the intra-modality information for learning better separate representation.
In the second learning stage, existing methods learn common representation with single-loss regularization \cite{feng12014cross}, which ignore the intrinsic relevance of intra-modality and inter-modality correlation.
Besides, existing methods \cite{feng12014cross,DBLP:conf/ijcai/PengHQ16} only extract separate representation from the original instances, but ignore the rich complementary fine-grained clues provided by their patches. 
The great progress of fine-grained image classification \cite{DBLP:conf/aaai/HeP17} shows the effectiveness of modeling fine-grained patches that contain discriminative local parts in single-modal scenario. It can also make great contribution to cross-modal retrieval because fine-grained correlation between patches of different modalities can provide more precise and complementary cross-modal correlation to the original instances.

%Although some works such as \cite{DBLP:conf/mm/JiangWLZLTZ15,DBLP:journals/tcsv/PengZZH16} adopt the idea of patch segmentation to exploit the fine-grained information, they have limited performance because they adopt the traditional framework with linear projection, which is hard to model the complex cross-modal correlation with high non-linearity. 

%\begin{figure}[!t]
%	\centering
%	\includegraphics[width=0.4\textwidth]{Figure/CommonSpace_v8-eps-converted-to.pdf}
%	\caption{Illustrations of mainstream framework of the most existing cross-modal retrieval methods, which attempt to represent data of different modalities with the same ``feature'' type (namely common representation) in a common space, so that the similarity measure can be directly performed.
%	}
%	\label{fig_CommonSpace}
%\end{figure}

\begin{figure}[t]
	\centering
	\begin{minipage}[c]{1.0\linewidth}
		\centering
		\includegraphics[width=0.95\textwidth]{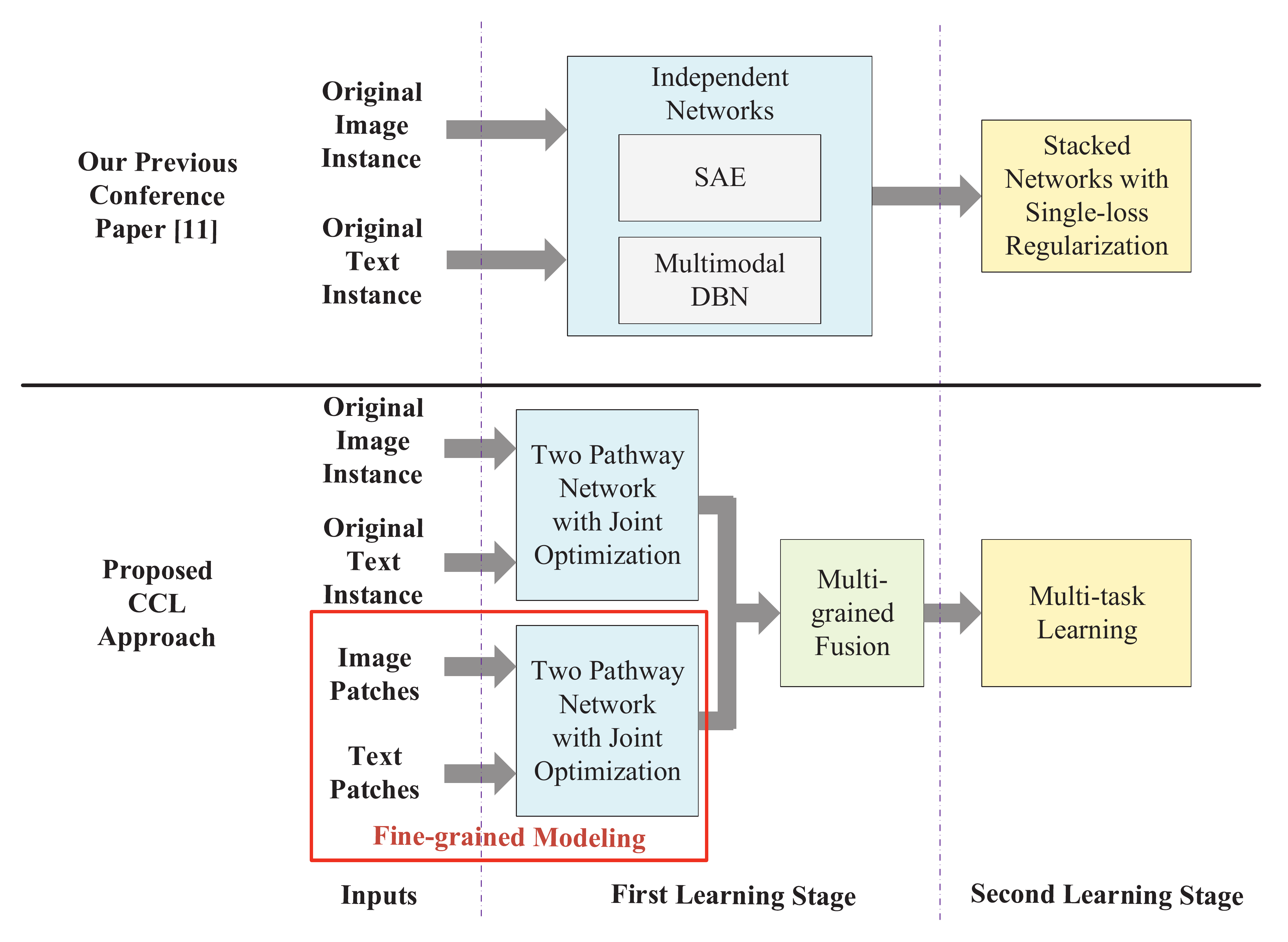}
	\end{minipage}%
	\caption{The schematic diagram to show the difference between the proposed CCL approach and our previous conference paper \cite{DBLP:conf/ijcai/PengHQ16}.}
	\label{fig:sch}
\end{figure}

For addressing the above issues, this paper proposes a cross-modal correlation learning (CCL) approach with multi-grained fusion by hierarchical network. Its main advantages and contributions can be summarized as follows:
\begin{itemize}
\item{
	%In the two-stage learning framework adopted by the existing methods, \textit{the first stage} is to generate the separate representation for each modality, and \textit{the second stage} is to get the cross-modal common representation. However,
	\textit{Cross-modal correlation exploiting. }
	In \textit{the first learning stage} for separate representation of each modality, existing methods \cite{feng12014cross,srivastava2012learning} only model intra-modality correlation but ignore inter-modality correlation. Actually, inter-modality correlation can provide rich complementary context to intra-modality one for learning better separate representation. So we employ multi-level association with joint optimization by maximizing intra-modality and inter-modality correlation simultaneously, which can capture the important hints from cross-modal correlation to boost common representation learning.
	} 
\item{
	\textit{Multi-task learning. }
	In \textit{the second learning stage} for common representation of different modalities, existing methods only adopt shallow networks with single-loss regularization \cite{DBLP:conf/icml/AndrewABL13,DBLP:conf/cvpr/YanM15}, 
	%which cannot effectively balance intra-modality and inter-modality correlation. 
	which ignore the intrinsic relevance of intra-modality and inter-modality correlation, so cannot effectively exploit and balance them to improve generalization performance. 
	%While the relative importance between them is not stable and needs adaptive tuning.
	%The semantic discriminative information within each modality has fluid relationship with the semantic correlation between different modalities, which needs adaptive tuning. 
	%The relative importance between the semantic discriminative information within intra-modality and semantic correlation within inter-modality is not stable and needs adaptive tuning.
	So we design a multi-task learning strategy to adaptively balance intra-modality semantic category constraints and inter-modality pairwise similarity constraints, 
	%which leads to more accurate representation for cross-modal common space.
	and make them mutually boost each other by fully exploiting their intrinsic relevance.
	} 
\item {
	\textit{Multi-grained fusion. }
	The patches contain complementary fine-grained clues to the original instances, which are ignored by the existing methods \cite{DBLP:conf/ijcai/PengHQ16,ngiam32011multimodal}. So we construct a multi-pathway network to fuse the multi-grained information in parallel by modeling the joint distributions, which can exploit and integrate the coarse-grained instances and fine-grained patches to make cross-modal correlation more precise.
	}
\end{itemize}
The main differences between the proposed CCL approach and our previous conference paper CMDN \cite{DBLP:conf/ijcai/PengHQ16} can be summarized as the following three aspects: 
(1) Our proposed CCL approach jointly employs the coarse-grained instances and fine-grained patches for \textit{multi-grained fusion} to learn more precise cross-modal correlation and boost cross-modal retrieval. While CMDN only uses the \textit{original coarse-grained instance}s, which ignores complementary fine-grained clues provided by their patches.
(2) Our proposed CCL approach adopts a \textit{multi-task learning strategy} to adaptively balance intra-modality semantic category constraints and inter-modality pairwise similarity constraints. While CMDN only adopts \textit{single-loss regularization}, which cannot effectively exploit and balance the above constraints to improve generalization performance.
(3) Our proposed CCL approach learns the separate representation in the first learning stage through \textit{one linked two-pathway network} by \textit{jointly optimizing} the intrinsic intra-modality and inter-modality correlation, which can fully capture these complementary hints simultaneously from cross-modal correlation. While CMDN learns the intra-modality and inter-modality separate representations \textit{respectively} by \textit{two independent networks}, which cannot effectively exploit the intrinsic relationship between these two kinds of complementary information. 
%Our previous conference paper \cite{DBLP:conf/ijcai/PengHQ16} adopts \textit{two independent networks} to learn intra-modality and inter-modality separate representation \textit{respectively}, %for the intra-modality and inter-modality, 
%CCL further models the \textit{two kinds of complementary information in a unified network with multi-level association by joint optimization}, 
%while our CCL approach in this paper further exploits the intrinsic intra-modality and inter-modality correlation with \textit{joint optimization in hierarchical network to learn multi-grained representation simultaneously}.
%which can fully exploit the intrinsic cross-modal correlation to boost the retrieval accuracy. 
A schematic diagram in Figure \ref{fig:sch} intuitively demonstrates the differences between CCL and CMDN.
To the best of our knowledge, the proposed CCL approach is the first to simultaneously model intra-modality and inter-modality correlation in both two learning stages, and employ the coarse-grained instances and fine-grained patches, which can learn more precise cross-modal correlation. Comparing with 13 state-of-the-art methods on 6 widely-used datasets, the effectiveness of our CCL approach is verified from the comprehensive experimental results.

The rest of this paper is organized as follows: Related works on cross-modal retrieval are briefly reviewed in Section II. Section III presents our proposed CCL approach. Section IV introduces the experiments as well as the results analysis. Finally Section V concludes this paper.

%\begin{figure*}[t]
%	\centering
%	\begin{minipage}[c]{1.0\linewidth}
%		\centering
%		\includegraphics[width=0.85\textwidth]{Figure/mDBN_BAE_CorrAE_v3-eps-converted-to.pdf}
%	\end{minipage}%
%	%\setlength{\abovecaptionskip}{0.2cm}
%	\caption{The network structures of Multimodal DBN, Bimodal Autoencoders and Correspondence Autoencoder. Multimodal DBN consists of two separate DBNs and a joint RBM on the top of them. Bimodal AE is a deep autoencoder network with a joint layer in the middle. And Correspondence Autoencoder has two subnetworks which are linked at the code layer.}
%	\label{fig:relatedNet}
%\end{figure*}

\section{Related Works}

In this section, we briefly review the representative methods of cross-modal retrieval, which are divided into two categories: Traditional methods and DNN-based methods.

\subsection{Traditional Cross-modal Retrieval Methods}

As for the traditional cross-modal retrieval methods, one representative method is Canonical Correlation Analysis (CCA) \cite{HotelingBiometrika36RelationBetweenTwoVariates}. Given a set of pairwise cross-modal data (such as image/text pairs), CCA learns a common space where the two  modalities  have maximum correlation.  Mapping matrices can be learned by CCA to project the features of different modalities into a lower-dimensional common space, and then common representation is obtained. CCA is widely used to model the multimodal data \cite{DBLP:journals/neco/HardoonSS04,BredinICASSP07CCAAudioVisual,DBLP:conf/cvpr/KleinLSW15} and also has many extensions and varieties \cite{RasiwasiaMM10SemanticCCA,DBLP:journals/ijcv/GongKIL14}.
For example, Rasiwasia et al. \cite{RasiwasiaMM10SemanticCCA} attempt to combine category information with CCA, and Multi-view CCA \cite{ DBLP:journals/ijcv/GongKIL14} extends CCA with the third view of high-level semantics. Similar to CCA, Li et al. \cite{LiMM03CFA} propose Cross-modal Factor Analysis (CFA) which also models the pairwise cross-modal correlation, but learns the projections by minimizing the Frobenius norm between pairwise data in common space. 
Ranjan et al. \cite{DBLP:conf/iccv/RanjanRJ15} propose multi-label CCA as an extension of CCA, which does not rely on the pairwise correspondence but considers the high-level semantic information in the form of multi-label annotations. Tran et al. \cite{DBLP:conf/cvpr/TranBC16} embed the projections of visual and textual features into a local context that reflects the data distribution in the common space. Besides, %Hua et al. \cite{DBLP:journals/tmm/HuaWLCH16} propose a cross-modal correlation method with adaptive hierarchical semantic aggregation.
Hua et al. \cite{DBLP:journals/tmm/HuaWLCH16} propose a cross-modal correlation method with adaptive hierarchical semantic aggregation, which constructs a set of local projections and probabilistic membership functions for image and text.

%Then a representation completion method is further proposed to deal with the situation when one modality is missing. And Rashtchian et al. \cite{rashtchian2010collecting} explore the image annotation by using Mechanical Turk and improve its quality by the use of a qualification test. And they construct large-scale image corpora with descriptive captions.

More recently, some methods apply semi-supervised learning and graph regularization into cross-modal common representation learning. For example, Joint Graph Regularized Heterogeneous Metric Learning (JGRHML) \cite{ZhaiAAAI2013JGRHML} proposed by Zhai et al. adopt metric learning and graph regularization to learn the project matrices, which constructs a joint graph regularization term using the data in the learned metric space. %further improves the former work by using semi-supervised information and simultaneously learns project matrices by constructing a separate graph for each modality. 
Joint Representation Learning (JRL) \cite{ZhaiTCSVT2014JRL} is proposed to construct a separate graph for each modality to learn a common space, which uses semantic information with semi-supervised regularization and sparse regularization.
%Also some works such as \cite{DBLP:journals/tmm/ZhuangYW08,DBLP:journals/tmm/WuYYTZZ14,yang2008harmonizing} model multimodal correlation by a uniform graph.
%Zhuang et al. \cite{DBLP:journals/tmm/ZhuangYW08} also construct a uniform cross-modal correlation graph to perform similarity propagation on multimodal data. 
%Besides, a two-level graph construction strategy is proposed by Yang et al. \cite{yang2008harmonizing} to construct a semantic space for performing cross-modal retrieval.
Wang et al. \cite{DBLP:journals/pami/WangHWWT16} adopt multimodal graph regularization term on the projected data with an iterative algorithm, which aims to preserve inter-modality and intra-modality similarity relationships.

\subsection{DNN-based Cross-modal Retrieval Methods}

%However, the performance of these methods mentioned above is limited for their traditional framework with linear projections. 
Deep learning has shown its strong power in modeling non-linear correlation, and achieved state-of-the-art performance in some applications of single-modal scenario, such as object detection \cite{DBLP:conf/cvpr/HeZRS16,DBLP:conf/nips/DaiLHS16} and image/video classification \cite{DBLP:conf/nips/KrizhevskySH12,DBLP:conf/mm/WuJWYX16}. Inspired by this, researchers attempt to model the complex cross-modal correlation with DNN, and the existing methods can be divided into two learning stages. \textit{The first learning stage} is to generate separate representation for each modality. And \textit{the second learning stage} is to learn common representation, which is the main focus of most existing methods based on DNN \cite{feng12014cross,srivastava2012learning}.
%Ngiam et al. \cite{ngiam32011multimodal} extend Restricted Boltzmann Machine (RBM) to generate common representation. Following this idea, similar network structures are proposed as %\cite{zhang2014start,kim2012learning,srivastava42012multimodal,DBLP:conf/ijcai/WangCO015}. 
%\cite{DBLP:journals/tmm/WangCO015,DBLP:journals/tmm/PangZN15,DBLP:conf/ijcai/WangCO015}. 
%Besides, some methods attempt to combine DNN with CCA to form Deep Canonical Correlation Analysis (DCCA) \cite{DBLP:conf/icml/AndrewABL13,DBLP:conf/cvpr/YanM15}, which maximizes the correlation on the top of two subnetworks.  Corr-AE \cite{feng12014cross} models the reconstruction error and correlation loss. %More recently, Peng et al.  \cite{DBLP:conf/ijcai/PengHQ16} propose a hierarchical multiple network structure for better retrieval accuracy. 
We briefly introduce some representative cross-modal retrieval methods based on DNN as follows:

\textbf{The Multimodal Deep Belief Network (Multimodal DBN)} \cite{srivastava2012learning} is proposed to learn common representation for the data of different modalities. %As shown in Figure \ref{fig:relatedNet}(a), 
In \textit{the first learning stage} for separate representation, it adopts a two-layer DBN for each modality to model the distribution of original features, where Gaussian Restricted Boltzmann Machine (RBM) is adopted for image instances, while Replicated Softmax model \cite{DBLP:conf/nips/SalakhutdinovH09} is used for text instances. RBM has several visible units $v$ and hidden units $h$, which is the basic component of DBN, and the energy function and joint distribution are defined as follows:
\begin{align}
&E(v,h;\theta )=-a^{\mathrm{T}}v-b^{\mathrm{T}}h-v^{\mathrm{T}}Wh \\
&P(v,h;\theta )= \frac{1}{Z(\theta)}exp(-E(v,h;\theta))
\end{align}
where $\theta$ is the collection of three parameters $a, b, W$ ($a, b$ are the bias parameters and $W$ is the weight parameter) and $Z(\theta)$ is the normalizing constant. Then in \textit{the second learning stage}, multimodal DBN applies a joint RBM on top of the two separate DBNs and combines them by modeling the joint distribution of data with different modalities to get common representation.

\textbf{The Bimodal Autoencoders (Bimodal AE)} \cite{ngiam32011multimodal} proposed by Ngiam et al. is based on deep autoencoder network, which is actually an extension of RBM for modeling multiple modalities. %As shown in Figure \ref{fig:relatedNet}(b), 
It has two subnetworks to learn separate representation in \textit{the first learning stage}, and then the two subnetworks are linked at the shared joint layer to generate common representation in \textit{the second learning stage}. Bimodal AE reconstructs different modalities such as image and text jointly by minimizing the reconstruction error between the original feature and reconstructed representation. 
Bimodal AE can learn high-order correlation between multiple modalities and preserve the reconstruction information at the same time.

\textbf{Correspondence Autoencoder (Corr-AE)} \cite{feng12014cross} first adopts DBN to generate separate representation in \textit{the first learning stage}. And then in \textit{the second learning stage}, it jointly models the correlation and reconstruction information with two subnetworks linked at the code layer, which minimizes a combination of representation learning error within each modality and correlation learning error between different modalities. 
Corr-AE, which only reconstructs the input itself, has two similar structures for extension: Corr-Cross-AE and Corr-Full-AE. Corr-Cross-AE attempts to reconstruct the input from different modalities, while Corr-Full-AE can reconstruct both the input itself and the input of different modalities.

\textbf{Cross-media Multiple Deep Networks (CMDN)} (our previous conference paper \cite{DBLP:conf/ijcai/PengHQ16} ) jointly models the complementary intra-modality and inter-modality correlation between different modalities in \textit{the first learning stage}. It should be noted that two independent networks are adopted in \textit{ the first learning stage} of CMDN. Specifically, Stacked Autoencoder (SAE) \cite{SAE} is used to model intra-modality correlation, while the Multimodal DBN is used to capture inter-modality correlation. In \textit{the second learning stage}, a hierarchical learning strategy is adopted to learn the cross-modal correlation with a two-level network, and common representation is learned by a stacked network based on Bimodal AE. 
The above DNN-based methods have three limitations in summary as follows.  

\textit{1)} In \textit{the first learning stage}, the existing methods as \cite{feng12014cross,srivastava2012learning} only model intra-modality correlation to generate separate representation, but ignore the rich complementary context provided by inter-modality correlation, which should be preserved for learning better separate representation. Although our previous work \cite{DBLP:conf/ijcai/PengHQ16} also considers intra-modality and inter-modality correlation in the first learning stage, it adopts two independent networks to model each of them respectively, which cannot fully exploit the complex relationship between intra-modality and inter-modality correlation.
While our proposed CCL approach models the two kinds of complementary information by jointly optimizing intra-modality reconstruction information and inter-modality pairwise similarity.

\textit{2)} In \textit{the second learning stage}, existing methods learn common representation by adopting shallow network architectures with single-loss regularization \cite{DBLP:conf/icml/AndrewABL13,DBLP:conf/cvpr/YanM15}. 
However, the intra-modality and inter-modality correlation has intrinsic relevance, and such relevance is ignored by the single-loss regularization, which leads to inability for improving generalization performance. 
Multi-task learning (MTL) framework has been proposed to enhance the generalization ability by constructing a series of learning processes, which are relevant to each other and can mutually boost each other.
%To enhance the generalization ability, researchers have proposed multi-task learning (MTL) framework to construct a series of parallel learning process. 
Recently, extensive research works attempt to apply multi-task learning into deep architecture. %, such as face recognition, image classification and so on.
DeepID2 \cite{Sun2014Deep} simultaneously learns face identification and verification as two learning tasks to achieve better accuracy of face recognition. Ren et al. \cite{DBLP:conf/nips/RenHGS15} propose Faster R-CNN, which also consists of two learning tasks as the object bound and objectness score prediction, and boosts the object detection accuracy. Besides, a joint multi-task learning algorithm \cite{DBLP:journals/tmm/AbdulnabiWLJ15} is proposed to predict attributes in images.
However, most of the research efforts have focused on the single-modal scenario. Inspired by the above methods, we apply multi-task learning to perform common representation learning. It aims to balance intra-modality semantic category constraints and inter-modality pairwise similarity constraints to further improve the accuracy of cross-modal retrieval.

\textit{3)} Furthermore, only the original instances are considered by the existing methods based on DNN \cite{feng12014cross,DBLP:conf/ijcai/PengHQ16,ngiam32011multimodal}. 
Although patches have been exploited in some traditional methods as \cite{DBLP:journals/tcsv/PengZZH16}, the accuracies of these methods are limited because of the traditional framework, which cannot effectively model the complex correlation between the patches with high non-linearity.
Our proposed CCL approach can fully exploit the coarse-grained instances as well as the rich complementary fine-grained patches by DNN, and fuse the multi-grained information to capture the intrinsic correlation between different modalities.
%Although some research efforts such as \cite{DBLP:journals/tcsv/PengZZH16} construct one hypergraph using the fine-grained information by adopting segmentation, their accuracies are limited by the traditional framework, which cannot effectively model the complex correlation between the patches. 
%Besides, C$^2$MLR \cite{DBLP:conf/mm/JiangWLZLTZ15} is proposed by Jiang et al. to perform cross-modal retrieval by optimizing a ranking function with both local and global alignment. 
%However, its main consideration is the ranking information, while we focus on exploiting the rich complementary fine-grained information in the original modality instances as well as their patches by DNN, and fully modeling the intrinsic semantic correlation between different modalities.

\begin{figure*}[t]
	\centering
	\begin{minipage}[c]{0.85\linewidth}
		\centering
		\includegraphics[width=0.9\textwidth]{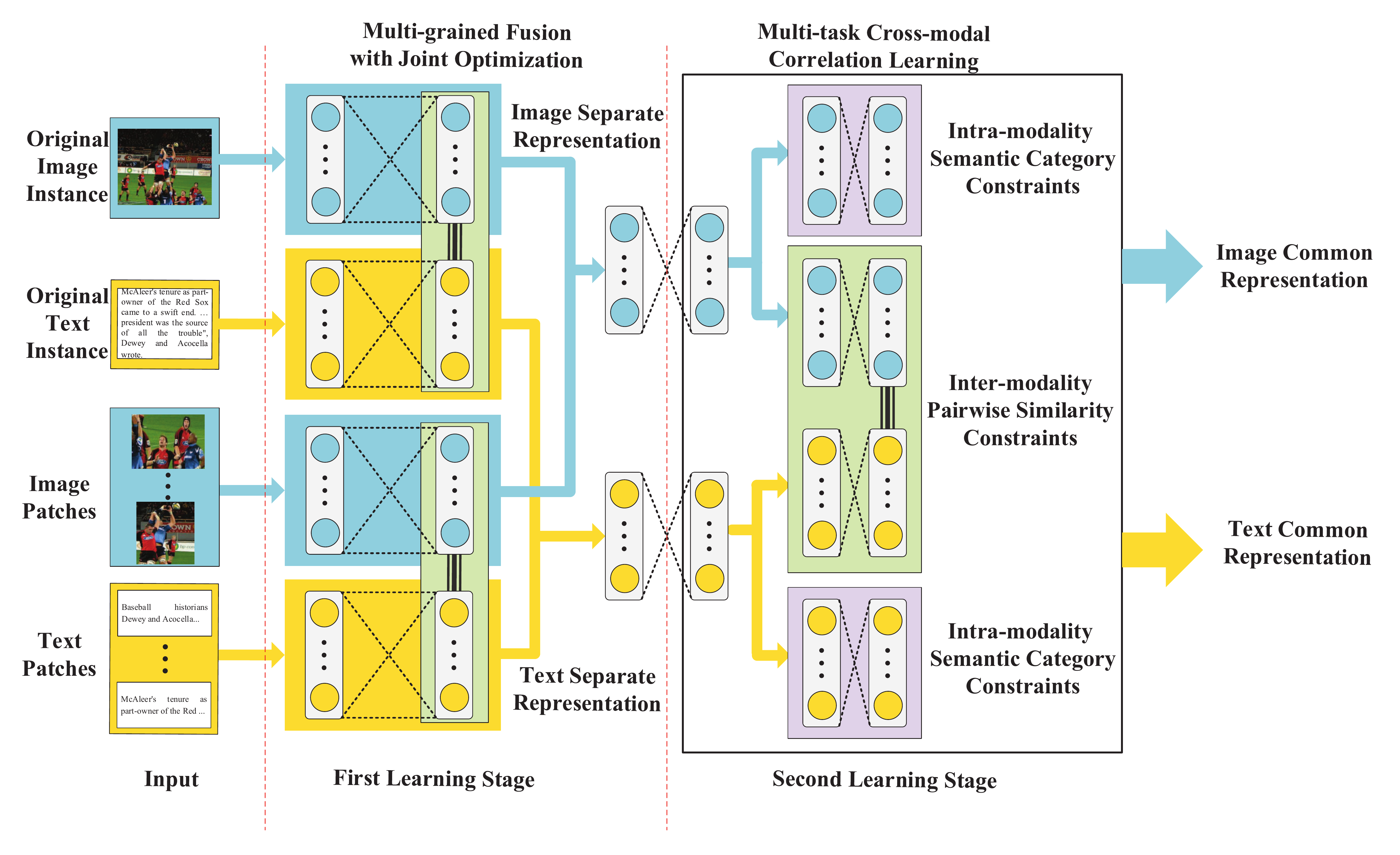}
	\end{minipage}%
	\caption{An overview of our CCL approach with two learning stages:
		In \textit{the first learning stage}, we learn separate representation by simultaneously modeling intra-modality and inter-modality correlation, which integrates the original instances and their patches in parallel. Then in \textit{the second learning stage}, we adopt a multi-task learning strategy to adaptively balance intra-modality and inter-modality correlation, which leads to more accurate common representation.}\label{fig:network}
\end{figure*}

\section{Our CCL Approach}

As shown in Figure \ref{fig:network}, CCL simultaneously models intra-modality and inter-modality correlation in both two learning stages, and employs the coarse-grained instances and fine-grained patches, which can learn more precise cross-modal correlation.
%our CCL model adopts a unified multi-pathway network with multi-level association to preserve the intra-modality information and inter-modality correlation simultaneously in both two learning stage. 
%In addition, in the first learning stage, we learn the separate representation by simultaneously modeling the original instances and their patches in parallel. Then in the second learning stage, we adopt a multi-task learning strategy to adaptively balance the information from intra-modality and inter-modality modeling, which leads to more accurate common representation.

%As shown in Figure \ref{fig:network}, our CCL model consists of two learning stages as follows. In the first learning stage, automatic patch segmentation method \cite{DBLP:journals/ijcv/UijlingsSGS13} is adopted on each modality, then we learn the separate representation for both the original modality instances and their patches to exploit the fine-grained semantic information, which also simultaneously models the intra-modality and inter-modality correlation at the same time.
%In the second learning stage, a two-level network structure is constructed including the joint RBM to combine the two types of separate representation for each modality, and a multi-task learning framework to learn the common representation.
%fine-grained separate representation learning and multi-task common representation learning, which is shown in Figure \ref{fig:network}.

The formal definition will first be given. The multimodal dataset consists of two modalities with $m$ image instances and $n$ text instances, which is denoted as $D=\left \{D^{(i)},D^{(t)}\right \}$. Here $D^{(i)}=\left \{x_{p}^{(i)},y_{p}^{(i)}\right \}_{p=1}^{m}$ denotes the data of image modality, where the $p$-th image instance is denoted as $x_{p}^{(i)}\in \mathbb{R}^{d^{(i)}}$ with its corresponding label $y_{p}^{(i)}$ and the dimensional number $d^{(i)}$. 
$D^{(t)}=\left \{x_{q}^{(t)},y_{q}^{(t)}\right \}_{q=1}^{n}$ denotes the data of text modality, where the text instance is defined as $x_{q}^{(t)}\in \mathbb{R}^{d^{(t)}}$ with the label $y_{q}^{(t)}$ and the dimensional number $d^{(t)}$. 
Besides, the pairwise correspondence is denoted as $(x_p^{(i)},x_p^{(t)})$, which means that the two instances of different modalities co-exist to describe the relevant semantics.
%In addition, the cross-modal retrieval task aims to retrieve the relevant text $x^{(t)}$ given the query of image $x^{(i)}$ in the unlabeled dataset of text $D_{u}^{(t)}=\left \{x_{1}^{(t)},...,x_{l}^{(t)}\right \}$ with $l$ text instances and vice-versa.
%The multimodal dataset is denoted as $D=\left \{D^{(i)},D^{(t)}\right \}$ with two types of modality, which has $m$ image instances and $n$ text instances. As for the image modality $D^{(i)}=\left \{x_{p}^{(i)},y_{p}^{(i)}\right \}_{p=1}^{m}$, the $p$-th image instance with its label $y_{p}^{(i)}$ is denoted as $x_{p}^{(i)}\in \mathbb{R}^{d^{(i)}}$, whose dimensional number is $d^{(i)}$. And text modality is defined similar as $D^{(t)}=\left \{x_{q}^{(t)},y_{q}^{(t)}\right \}_{q=1}^{n}$, where $x_{q}^{(t)}\in \mathbb{R}^{d^{(t)}}$ labeled as $y_{q}^{(t)}$. 

\subsection{First Learning Stage: Multi-grained Fusion with Joint Optimization}

In the first learning stage, we construct a multi-pathway network, which aims to obtain separate representations from both the original instances as well as their patches of each modality in parallel, and capture intra-modality and inter-modality correlation with joint optimization at the same time.% for the first level association.

\subsubsection{\textbf{Coarse-grained learning with original instances}} A two-pathway network structure is adopted to model the image and text instances. First, two types of Deep Belief Network (DBN) \cite{DBLP:journals/neco/HintonOT06} are used to model the distribution over the features of each modality, where Gaussian Restricted Boltzmann Machine (RBM) \cite{DBLP:conf/nips/WellingRH04} is adopted to model the image instances and Replicated Softmax model \cite{DBLP:conf/nips/SalakhutdinovH09} is adopted for text instances. We define the probability functions of each DBN as follows:

%We adopt relatively general deep models for the generality of cross-modal retrieval to form the multi-pathway network, and two types of Deep Belief Network (DBN) \cite{DBLP:journals/neco/HintonOT06} are used to model the distribution over the features of each modality, where Gaussian Restricted Boltzmann Machine (RBM) is adopted to model the image instances and Replicated Softmax model \cite{DBLP:conf/nips/SalakhutdinovH09} is for text instances.

%To consider the generality of cross-modal retrieval, relatively general deep models are adopted, which are two types of Deep Belief Network (DBN) \cite{DBLP:journals/neco/HintonOT06} to model the distribution over the features of each modality. Gaussian Restricted Boltzmann Machine (RBM) is used to model the image instances. And for text instances, we use Replicated Softmax model \cite{DBLP:conf/nips/SalakhutdinovH09}. 

%First, we attempt to obtain the separate representation from both the original modality instances as well as their patches of each modality. As for the original modality instances, two types of DBN as we mentioned above are adopted directly over the features of image and text instances. And we define the probability functions of each DBN as follows:
%First, we obtain the separate representation for each modality from both the original modality instances and their patches. For the original modality instances, we directly adopt two types of DBN on the features of image and text. And the probability functions of each DBN are defined as follows:
\begin{gather}
P(v_{i})=\sum_{h^{(1)},h^{(2)}}P(h^{(2)},h^{(1)})P(v_{i}|h^{(1)}) \\
P(v_{t})=\sum_{h^{(1)},h^{(2)}}P(h^{(2)},h^{(1)})P(v_{t}|h^{(1)})
\end{gather}
where the two hidden layers of DBN are denoted as $h^{(1)}$ and $h^{(2)}$, while $v_{i}$ is for image input and $v_{t}$ is for text input. The outputs of two DBNs can preserve the original characteristic of each modality with high-level semantic information, which are denoted as $Q^{(i)}$ and $Q^{(t)}$. 

Then %inspired by \cite{feng12014cross}, 
we simultaneously model intra-modality and inter-modality correlation by joint optimization for $Q^{(i)}$ of image instance and $Q^{(t)}$ of text instance. Compared with our previous CMDN method \cite{DBLP:conf/ijcai/PengHQ16}, which adopts two independent networks for intra-modality and inter-modality to learn separate representation, a two-pathway network linked at the top code layer is constructed. We minimize the following loss function to jointly optimize the reconstruction learning error and correlation learning error:
\begin{align}
L(Q^{(i)}, Q^{(t)})=&L_{r}(Q^{(i)},Q_{r}^{(i)})+L_{r}(Q^{(t)},Q_{r}^{(t)}) \notag \\
&+L_{c}(Q^{(i)},Q^{(t)})
\end{align}
\begin{align}
L_{r}(Q^{(k)},Q_{r}^{(k)})&=\left \| Q^{(k)}-Q_{r}^{(k)} \right \|^{2}, k=i,t \label{eq:lr}\\
L_{c}(Q^{(i)},Q^{(t)})&=\left \| Q^{(i)}-Q^{(t)} \right \|^{2} \label{eq:lc}
\end{align}
where $Q_{r}^{(i)}$ and $Q_{r}^{(t)}$ denote the reconstruction representations of image and text, and $L_{r}$ in Eq.(\ref{eq:lr}) represents the loss of reconstruction learning error, which aims to minimize the L2 distance between the input and reconstruction representation of each modality. $L_{c}$ in Eq.(\ref{eq:lc}) is for correlation learning error to minimize the L2 distance between the instances of different modalities. Thus, we can get the coarse-grained representations with both intra-modality and inter-modality correlation for the original instances of different modalities, which are denoted as $T_{origin}^{(i)}$ and $T_{origin}^{(t)}$. 

\subsubsection{\textbf{Fine-grained learning with patches}} 
\begin{figure}[!t]
	\centering
	\includegraphics[width=3.0in]{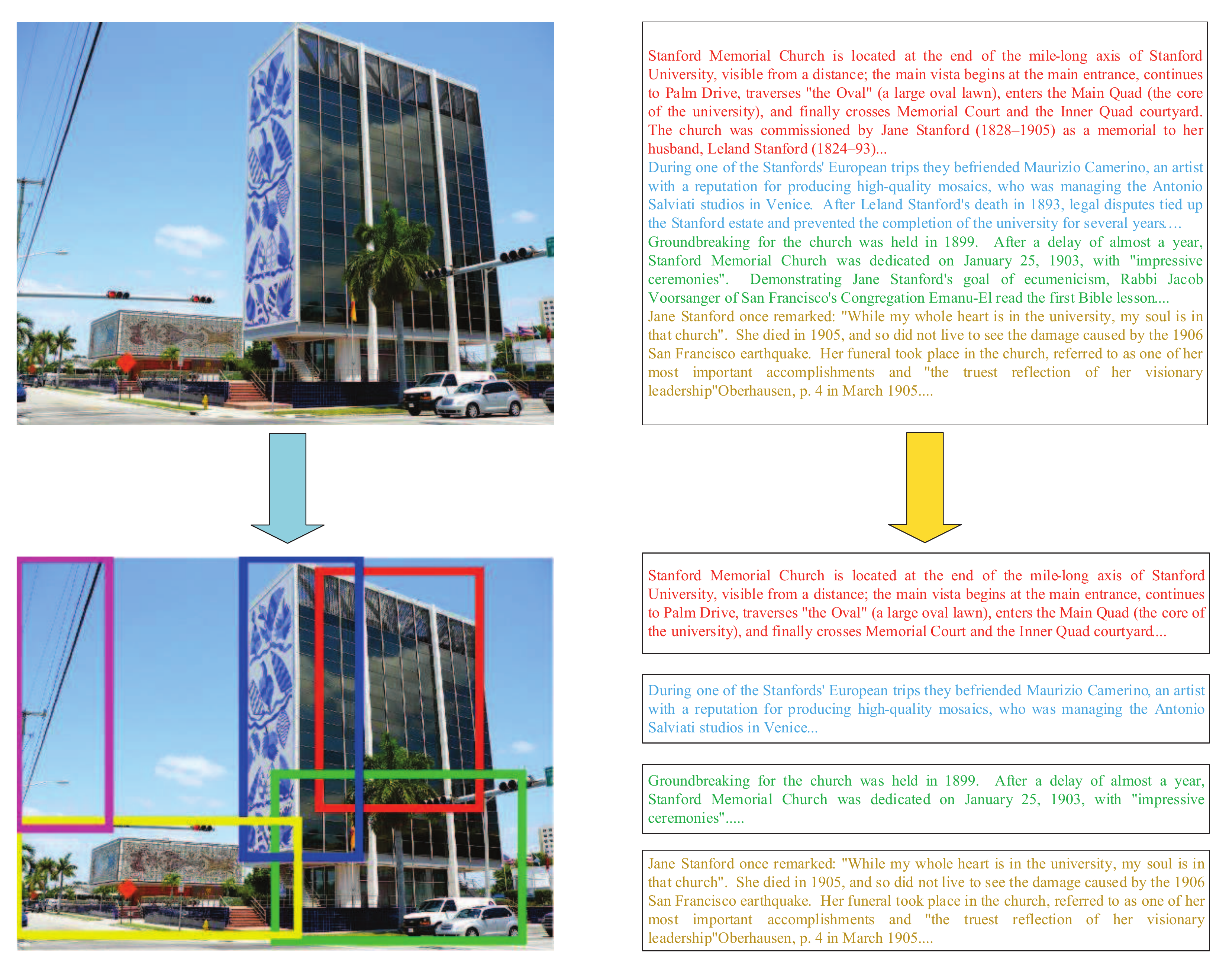}
	\caption{Examples for the generation of fine-grained patches.}
	\label{fig_seg}
\end{figure}

We first divide each original image and text instance into several patches, which is showed in Figure \ref{fig_seg}. Specifically, we adopt selective search \cite{DBLP:journals/ijcv/UijlingsSGS13} to extract region proposals, which can find the visual objects in the image instance containing rich fine-grained information. For text, the segmentation is performed according to the form of text, where the text is divided into paragraphs, sentences or words.
Similar with the original instances, a two-pathway network structure is constructed with two types of DBN adopted over the features extracted from the patches of image and text. For the patches within one original instance, average fusion is adopted to combine their representations obtained from DBN, and the results are denoted as $U^{(i)}$ and $U^{(t)}$. Then we link the two-pathway network at the code layer, and minimize the following loss function to model intra-modality and inter-modality correlation with joint optimization:
%As shown in Figure \ref{fig_seg}, for the patches of different modalities, we first divide the input image and text into several patches. Specifically, selective search \cite{DBLP:journals/ijcv/UijlingsSGS13} is adopted to extract several region proposals to find the visual objects in the image instance, which contain rich fine-grained semantic information. Similarly, two types of DBN are adopted on the features extracted from the patches of image and text, which are the same with the original modality instances. And we adopt average fusion to combine the representation for all patches within one instance, denoted as $U^{(i)}$ and $U^{(t)}$. Then, a similar two-pathway network structure with a linked code layer is constructed, and the following loss function is minimized to simultaneously model the intra-modality and inter-modality information.
\begin{align}
L(U^{(i)}, U^{(t)})=&L_{r}(U^{(i)},U_{r}^{(i)})+L_{r}(U^{(t)},U_{r}^{(t)}) \notag \\
&+L_{c}(U^{(i)},U^{(t)})
\end{align}
where the loss of reconstruction learning error is represented as $L_{r}$, and the loss of correlation learning error is denoted as $L_{c}$. They have similar definition with Eq.(\ref{eq:lr}) and Eq.(\ref{eq:lc}).
Therefore, the fine-grained representations denoted as $T_{patch}^{(i)}$ and $T_{patch}^{(t)}$ are obtained from patches of different modalities, which preserve fine-grained intra-modality and inter-modality correlation.
%By employing the first level association with multi-pathway network, the learned separate representation can capture import hint from cross-modal correlation as well as the complementary fine-grained clues to boost the common representation learning in the following learning stage.

\subsubsection{\textbf{Multi-grained Fusion}} 
%With the multi-pathway network mentioned above, we have obtained multiple complementary representation $T_{origin}^{(i)}$, $T_{patch}^{(i)}$ and $T_{origin}^{(t)}$, $T_{patch}^{(t)}$ for each modality, which captures multi-grained information as well as intra-modality and inter-modality correlation. 
We adopt a joint RBM for each modality to fuse the coarse-grained and fine-grained representations obtained from both the original instances ($T_{origin}^{(i)}$, $T_{origin}^{(t)}$) and their patches ($T_{patch}^{(i)}$, $T_{patch}^{(t)}$). And the joint distribution is defined as follows:
\begin{gather}
P(v_{1},v_{2})=\sum_{h_{1}^{(1)},h_{2}^{(1)},h^{(2)}}P(h_{1}^{(1)},h_{2}^{(1)},h^{(2)})\times \notag \\
\sum_{h_{1}^{(1)}}P(v_{1}\mid h_{1}^{(1)})\times \sum_{h_{2}^{(1)}}P(v_{2}\mid h_{2}^{(1)})
\end{gather}
where the two types of intermediate representation $T_{origin}^{(i)}$ and $T_{patch}^{(i)}$ for image instances are denoted as $v_{1}$ and $v_{2}$. Thus, this joint distribution is collected as separate representation for image, which can also be adopted on the two intermediate representations of text instances $T_{origin}^{(t)}$ and $T_{patch}^{(t)}$ to generate separate representation for text. The obtained separate representations for image and text are denoted as $S^{(i)}$ and $S^{(t)}$, which capture the intrinsic correlation and rich complementary information contained in both original instances and their patches of each modality.%, and are taken as the input of the second learning stage.

\subsection{Second Learning Stage: Multi-task Cross-modal Correlation Learning}

In the second learning stage for generating common representation, as shown in the right of Figure \ref{fig:network}, we propose a multi-task learning framework to model intra-modality semantic category constraints and inter-modality pairwise similarity constraints as two loss branches. The former aims to improve the semantic discriminative ability in the high-level common space, while the later can capture the intrinsic correlation between different modalities, which leads to more accurate common representation for cross-modal data. 

For inter-modality pairwise similarity constraints, most existing works as \cite{feng12014cross,DBLP:conf/ijcai/PengHQ16} only focus on pairwise correlation with similar constraints but ignore the semantically dissimilar constraints. Thus we model the pairwise similar and dissimilar constraints between different modalities with a contrastive loss, which has the following considerations: Image and text instances with the same label should be similar and the distance between them should be minimized, while on the contrary, image and text instances which have different labels should be dissimilar and their distance should be maximized. Specifically, a neighborhood graph $G=(V,E)$ is constructed in a mini-batch of data for one iteration, where the vertices $V$ represent the image and text instances, and $E$ is the similarity matrix between data of two modalities according to their labels, which is defined as follows:
%For the inter-modality correlation learning task, we adopt the contrastive loss to model the pairwise similar and dissimilar constraints between different modalities, with the following two considerations: the distance between the image and text instances with the same label should be minimized, while the distance between the image and text instances which have different labels should be maximized. It should be noted that we construct a neighborhood graph $G=(V,E)$ in a mini-batch of data for one iteration, where the image and text instances are represented as vertices $V$, and $E$ represents the similarity matrix between the data of two modalities according to their labels, defined as follows:
\begin{gather}
E(p,q)=\begin{cases}
1 & : y_{p}^{(i)}=y_{q}^{(t)}\\ 
0 & : y_{p}^{(i)}\neq y_{q}^{(t)} 
\end{cases}
\label{eq_e}
\end{gather}
Thus, the contrastive loss between the image and text pairs is defined to model the pairwise similar and dissimilar constraints as follows:
\begin{gather}
L_{1}(p,q)=\begin{cases}
\left \| f(s_{p}^{i})-g(s_{q}^{t}) \right \|^{2} & E(p,q)=1 \\ 
max(0,\alpha-\left \| f(s_{p}^{i})-g(s_{q}^{t}) \right \|^{2}) & E(p,q)=0 
\end{cases} \label{equ:ls}
\end{gather}
where $s_{p}^{i}$ and $s_{q}^{t}$ denote separate representations for image $S^{(i)}$ and text $S^{(t)}$, 
$f(.)$ and $g(.)$ denote the non-linear mappings respectively for image and text pathways in the multi-task learning network in the second learning stage. Each pathway consists of three fully-connected layers, aiming to convert the separate representations $S^{(i)}$ and $S^{(t)}$ to the final common representations of image and text.
The margin parameter is set to be $\alpha$. Then we calculate the derivative of loss function $L_{1}$ in Eq.(\ref{equ:ls}) for positive pair of image $s_{p}^{i}$ and text $s_{q}^{t+}$ where $E(p,q)=1$ as follows:
\begin{align}
\frac{\partial L_{1}}{\partial f(s_{p}^{i})}=2(f(s_{p}^{i})-g(s_{q}^{t+})) \\
\frac{\partial L_{1}}{\partial g(s_{q}^{t+})}=2(g(s_{q}^{t+})-f(s_{p}^{i}))
\end{align}
then for the negative pair of image $s_{p}^{i}$ and text $s_{q}^{t-}$ where $E(p,q)=0$, the derivative of loss function is calculated as follows:
\begin{align}
\frac{\partial L_{1}}{\partial f(s_{p}^{i})}=2(g(s_{p}^{i})-f(s_{q}^{t-})) \times J \\
\frac{\partial L_{1}}{\partial g(s_{q}^{t-})}=2(f(s_{q}^{t-})-g(s_{p}^{i})) \times J
\end{align}
%where the image intermediate representation is denoted as $s_{p}^{i}$, $s_{q}^{t}$ denotes the text intermediate representation, and $\alpha$ is the margin parameter. And the derivative of the loss function $L_{1}$ in (\ref{equ:ls}) is calculated for each image $p$ and text $q$ as follows:
%\begin{equation}
%\begin{split}
%& \frac{\partial L_{1}}{\partial f(s_{p}^{i})}=\sum_{q=1,E(p,q)=1}^{m}J(s_{p}^{i},s_{q}^{t}) \\
%& -\sum_{q=1,E(p,q)=0}^{m}(\alpha-\left \| J(s_{p}^{i},s_{q}^{t}) \right \|)\times sgn(J(s_{p}^{i},s_{q}^{t}))\\
%\end{split}
%\end{equation}
%\begin{equation}
%\begin{split}
%& \frac{\partial L_{1}}{\partial g(s_{q}^{t})}=\sum_{p=1,E(p,q)=1}^{n}J(s_{q}^{t},s_{p}^{i}) \\
%& -\sum_{p=1,E(p,q)=0}^{n}(\alpha-\left \| J(s_{q}^{t},s_{p}^{i}) \right \|)\times sgn(J(s_{q}^{t},s_{p}^{i})))\\
%\end{split}
%\end{equation}
where $J=0$ if $\alpha-\left \| f(s_{p}^{i})-g(s_{q}^{t-}) \right \|^{2} <= 0$, otherwise $J$ is set to be $1$. Therefore, the back-propagation can be applied to update the parameters through the network.

Then, for intra-modality semantic category constraints, a classification process is employed to exploit the intrinsic semantic information within each modality, which can classify data of each modality into one of $n$  categories. Thus, we present intra-modality semantic category constraints as an $n$-way softmax layer, where $n$ is the number of categories. Cross-entropy loss is minimized as follows:
%Then, for the intra-modality identification task, we employ a classification process to classify the data of each modality into one of $n$ different identities, where $n$ is set according to the class number of dataset. This identification task is presented with an $n$-way softmax layer to capture the semantic correlation within each modality. Thus, we minimize the cross-entropy loss as follows:
\begin{gather}
L_{2}=-\sum_{i=1}^{n}p_{i}log(\hat{p_{i}})
\end{gather}
where the predicted probability distribution is denoted as $\hat{p_{i}}$, and $p_{i}$ is the target probability distribution. By minimizing the above loss function, the semantical discrimination ability of common representation can be greatly enhanced.
%where $\hat{p_{i}}$ is the predicted probability distribution, while the target probability distribution is denoted as $p_{i}$. By minimizing the above loss function, it can enhance the semantically discrimination ability of the common representation.

%Finally, the common representations obtained from the multi-task learning framework, which are denoted as $M^{(i)}$ and $M^{(t)}$, can fully exploit the complex cross-modal correlation for boosting the accuracy of cross-modal retrieval.

Finally, we can obtain the accurate common representations denoted as $M^{(i)}$ and $M^{(t)}$ from the outputs of last fully-connected layer in the multi-task learning network, which can adaptively balance intra-modality semantic category constraints and inter-modality pairwise similarity constraints, and further make them mutually boost each other.
%Given a query of text instance, cross-modal retrieval aims to retrieve the relevant image instances in the unlabeled test set of image, and vice-versa. 
For performing the cross-modal retrieval, common representations of each image and text are extracted firstly from the above network structure with the inputs of both original instance and patches, and then the traditional distance metric, such as cosine distance, can be applied to measure the similarity between the instances of different modalities.

\begin{table*}[htb]
	\caption{The MAP scores of \textbf{Bi-modal Retrieval} for our CCL approach and compared methods on \textbf{Wikipedia} dataset. MAP with two types of image feature are calculated, which are \textbf{hand-crafted feature} and \textbf{CNN feature}, while the text feature is the same.}
	\begin{center}
		\scalebox{0.9}{
			\begin{tabular}{|c|c|c|c|c|c|c|} 
				\hline
				\multirow{2}{*}{Method} & \multicolumn{3}{c|}{MAP with hand-crafted feature} & \multicolumn{3}{c|}{MAP with CNN feature}\\
				\cline{2-7}
				& Image$\rightarrow$Text & Text$\rightarrow$Image & Average & Image$\rightarrow$Text & Text$\rightarrow$Image & Average\\
				\hline
				\textbf{Our CCL Approach} & \textbf{0.418} & \textbf{0.359} & \textbf{0.389} & \textbf{0.504} & \textbf{0.457} & \textbf{0.481}\\
				CMDN \cite{DBLP:conf/ijcai/PengHQ16} & 0.393 & 0.325 & 0.359 &0.488 &0.427 &0.458 \\
				LGCFL \cite{DBLP:journals/tmm/KangXLXP15} & 0.385 & 0.326 & 0.356 &0.481 &0.418 &0.450\\
				JRL \cite{ZhaiTCSVT2014JRL} & 0.344 & 0.277& 0.311 &0.453 &0.400 &0.427\\
				DCCA \cite{DBLP:conf/cvpr/YanM15} & 0.248 & 0.221& 0.235 &0.440 &0.390 &0.415\\
				Corr-AE \cite{feng12014cross} & 0.280 & 0.242& 0.261 &0.402 &0.395 &0.399\\
				Multimodal DBN \cite{srivastava2012learning} & 0.149 & 0.150& 0.150 &0.204 &0.183 &0.194\\
				Bimodal AE \cite{ngiam32011multimodal} & 0.236 & 0.208& 0.222 &0.314 &0.290 &0.302\\
				GMM+HGLMM \cite{DBLP:conf/cvpr/KleinLSW15} & 0.266 & 0.233& 0.250 &0.428 &0.396 &0.412\\
				Multi-label CCA \cite{DBLP:conf/iccv/RanjanRJ15} & 0.317 & 0.266& 0.292 &0.404 &0.366 &0.385\\
				MACC \cite{DBLP:conf/cvpr/TranBC16} & 0.255 & 0.233& 0.244 &0.470 &0.400 &0.435\\
				KCCA(Gaussian) \cite{DBLP:journals/neco/HardoonSS04} & 0.245 & 0.219& 0.232 &0.326 &0.268 &0.297\\
				KCCA(Poly) \cite{DBLP:journals/neco/HardoonSS04} & 0.200 & 0.185& 0.193 &0.215 &0.214 &0.215\\
				CFA \cite{LiMM03CFA} & 0.236 & 0.211& 0.224 & 0.334 &0.297 &0.316\\
				CCA \cite{HotelingBiometrika36RelationBetweenTwoVariates} & 0.203 & 0.183& 0.193 & 0.258 &0.250 &0.254\\
				\hline
				
			\end{tabular} 
		}
	\end{center}
	\vspace{-0.5cm}
	\label{table:wiki}
\end{table*}

\begin{table*}[htb]
	\caption{The MAP scores of \textbf{All-modal Retrieval} for our CCL approach and compared methods on \textbf{Wikipedia} dataset. MAP with two types of image feature are calculated, which are \textbf{hand-crafted feature} and \textbf{CNN feature}, while the text feature is the same.}
	\begin{center}
		\scalebox{0.9}{
			\begin{tabular}{|c|c|c|c|c|c|c|} 
				\hline
				\multirow{2}{*}{Method} & \multicolumn{3}{c|}{MAP with hand-crafted feature} & \multicolumn{3}{c|}{MAP with CNN feature}\\
				\cline{2-7}
				& Image$\rightarrow$All & Text$\rightarrow$All & Average & Image$\rightarrow$All & Text$\rightarrow$All & Average \\
				\hline
				\textbf{Our CCL Approach} & \textbf{0.331} & \textbf{0.610} & \textbf{0.471} & \textbf{0.422} & \textbf{0.652} & \textbf{0.537} \\
				CMDN \cite{DBLP:conf/ijcai/PengHQ16} & 0.282 & 0.592 & 0.437 &0.361 &0.637 &0.499 \\
				LGCFL \cite{DBLP:journals/tmm/KangXLXP15} & 0.294 & 0.576 & 0.435 &0.402 &0.601 &0.502\\
				JRL \cite{ZhaiTCSVT2014JRL} & 0.281 & 0.556& 0.419 &0.381 &0.530 &0.456\\
				DCCA \cite{DBLP:conf/cvpr/YanM15} & 0.214 & 0.317& 0.266 &0.374 &0.552 &0.463\\
				Corr-AE \cite{feng12014cross} & 0.225 & 0.401& 0.313 &0.311 &0.537 &0.424\\
				Multimodal DBN \cite{srivastava2012learning} & 0.140 & 0.177& 0.159 &0.170 &0.190 &0.180\\
				Bimodal AE \cite{ngiam32011multimodal} & 0.175 & 0.422& 0.299 &0.281 &0.517 &0.399\\
				GMM+HGLMM \cite{DBLP:conf/cvpr/KleinLSW15} & 0.225 & 0.335& 0.280 &0.362 &0.562 &0.462\\
				Multi-label CCA \cite{DBLP:conf/iccv/RanjanRJ15} & 0.253 & 0.484& 0.369 &0.333 &0.529 &0.431\\
				MACC \cite{DBLP:conf/cvpr/TranBC16} & 0.200 & 0.387& 0.294 &0.393 &0.557 &0.475\\
				KCCA(Gaussian) \cite{DBLP:journals/neco/HardoonSS04} & 0.163 & 0.377& 0.270 &0.321 &0.472 &0.397\\
				KCCA(Poly) \cite{DBLP:journals/neco/HardoonSS04} & 0.158 & 0.317& 0.238 &0.256 &0.320 &0.288\\
				CFA \cite{LiMM03CFA} & 0.174 & 0.283& 0.229 &0.300 &0.364 &0.332\\
				CCA \cite{HotelingBiometrika36RelationBetweenTwoVariates} & 0.180 & 0.315& 0.248 &0.219 &0.343 &0.281 \\
				\hline
				
			\end{tabular} 
		}
	\end{center}
	
	\label{table:wiki_all}
\end{table*}

\begin{table*}[htb]
	\caption{The MAP scores of \textbf{Bi-modal Retrieval} for our CCL approach and the compared methods on \textbf{NUS-WIDE-10K} dataset. We test two types of image feature: \textbf{Hand-crafted feature} and \textbf{CNN feature}, while the text feature is the same.}
	\begin{center}
		\scalebox{0.9}{
			\begin{tabular}{|c|c|c|c|c|c|c|} 
				\hline
				\multirow{2}{*}{Method} & \multicolumn{3}{c|}{MAP with hand-crafted feature} & \multicolumn{3}{c|}{MAP with CNN feature}\\
				\cline{2-7}
				& Image$\rightarrow$Text & Text$\rightarrow$Image & Average & Image$\rightarrow$Text & Text$\rightarrow$Image & Average \\
				\hline
				\textbf{Our CCL Approach} & \textbf{0.400} & \textbf{0.401} & \textbf{0.401} & \textbf{0.506} & \textbf{0.535} & \textbf{0.521} \\
				CMDN \cite{DBLP:conf/ijcai/PengHQ16} & 0.391 & 0.357 & 0.374 &0.492 &0.515 &0.504 \\
				LGCFL \cite{DBLP:journals/tmm/KangXLXP15} & 0.319 & 0.324 & 0.321 &0.428 &0.466 &0.447\\
				JRL \cite{ZhaiTCSVT2014JRL} & 0.324 & 0.263& 0.294 &0.426 &0.376 &0.401\\
				DCCA \cite{DBLP:conf/cvpr/YanM15} & 0.219 & 0.210& 0.215 &0.407 &0.416 &0.412\\
				Corr-AE \cite{feng12014cross} & 0.223 & 0.227& 0.225 &0.366 &0.417 &0.392\\
				Multimodal DBN \cite{srivastava2012learning} & 0.158 & 0.130& 0.144 &0.201 &0.259 &0.230\\
				Bimodal AE \cite{ngiam32011multimodal} & 0.159 & 0.172& 0.166 &0.327 &0.369 &0.348\\
				GMM+HGLMM \cite{DBLP:conf/cvpr/KleinLSW15} & 0.244 & 0.234& 0.239 &0.440 &0.453 &0.447\\
				Multi-label CCA \cite{DBLP:conf/iccv/RanjanRJ15} & 0.299 & 0.289& 0.294 &0.413 &0.437 &0.425\\
				MACC \cite{DBLP:conf/cvpr/TranBC16} & 0.167 & 0.157& 0.162 &0.453 &0.497 &0.475\\
				KCCA(Gaussian) \cite{DBLP:journals/neco/HardoonSS04} & 0.232 & 0.213& 0.223 &0.300 &0.336 &0.318\\
				KCCA(Poly) \cite{DBLP:journals/neco/HardoonSS04} & 0.150 & 0.149& 0.150 &0.114 &0.130 &0.122\\
				CFA \cite{LiMM03CFA} & 0.211 & 0.188& 0.200 &0.400 &0.299 &0.350\\
				CCA \cite{HotelingBiometrika36RelationBetweenTwoVariates} & 0.141 & 0.138& 0.140 &0.202 &0.220 &0.211 \\
				\hline
				
			\end{tabular} 
		}
	\end{center}
	\vspace{-0.5cm}
	\label{table:nus}
\end{table*}

\begin{table*}[htb]
	\caption{The MAP scores of \textbf{All-modal Retrieval} for our CCL approach and the compared methods on \textbf{NUS-WIDE-10K} dataset. We test two types of image feature: \textbf{Hand-crafted feature} and \textbf{CNN feature}, while the text feature is the same.}
	\begin{center}
		\scalebox{0.9}{
			\begin{tabular}{|c|c|c|c|c|c|c|} 
				\hline
				\multirow{2}{*}{Method} & \multicolumn{3}{c|}{MAP with hand-crafted feature} & \multicolumn{3}{c|}{MAP with CNN feature}\\
				\cline{2-7}
				& Image$\rightarrow$All & Text$\rightarrow$All & Average & Image$\rightarrow$All & Text$\rightarrow$All & Average \\
				\hline
				\textbf{Our CCL Approach} & \textbf{0.379} & \textbf{0.444} & \textbf{0.412} & \textbf{0.537} & \textbf{0.502} & \textbf{0.520} \\
				CMDN \cite{DBLP:conf/ijcai/PengHQ16} & 0.306 & 0.417 & 0.362 &0.478 &0.449 &0.464 \\
				LGCFL \cite{DBLP:journals/tmm/KangXLXP15} & 0.291 & 0.394 & 0.343 &0.434 &0.459 &0.447\\
				JRL \cite{ZhaiTCSVT2014JRL} & 0.237 & 0.421& 0.329 &0.445 &0.357 &0.401\\
				DCCA \cite{DBLP:conf/cvpr/YanM15} & 0.213 & 0.236& 0.225 &0.423 &0.405 &0.414\\
				Corr-AE \cite{feng12014cross} & 0.222 & 0.245& 0.234 &0.389 &0.379 &0.384\\
				Multimodal DBN \cite{srivastava2012learning} & 0.128 & 0.171& 0.150 &0.193 &0.338 &0.266\\
				Bimodal AE \cite{ngiam32011multimodal} & 0.145 & 0.257& 0.201 &0.255 &0.287 &0.271\\
				GMM+HGLMM \cite{DBLP:conf/cvpr/KleinLSW15} & 0.235 & 0.265& 0.250 &0.449 &0.443 &0.446\\
				Multi-label CCA \cite{DBLP:conf/iccv/RanjanRJ15} & 0.265 & 0.353& 0.309 &0.430 &0.413 &0.422\\
				MACC \cite{DBLP:conf/cvpr/TranBC16} & 0.159 & 0.171& 0.165 &0.418 &0.484 &0.451\\
				KCCA(Gaussian) \cite{DBLP:journals/neco/HardoonSS04} & 0.147 & 0.282& 0.215 &0.386 &0.351 &0.369\\
				KCCA(Poly) \cite{DBLP:journals/neco/HardoonSS04} & 0.138 & 0.173& 0.156 &0.304 &0.150 &0.227\\
				CFA \cite{LiMM03CFA} & 0.169 & 0.235& 0.202 &0.383 &0.314 &0.349\\
				CCA \cite{HotelingBiometrika36RelationBetweenTwoVariates} & 0.143 & 0.176& 0.160 &0.215 &0.216 &0.216 \\
				\hline
				
			\end{tabular} 
		}
	\end{center}
	%\vspace{-0.5cm}
	\label{table:nus_all}
\end{table*}

\begin{table*}[htb]
	\caption{The MAP scores of \textbf{Bi-modal Retrieval} for our CCL approach and the compared methods on \textbf{Pascal Sentence} dataset. We test two types of image feature: \textbf{Hand-crafted feature} and \textbf{CNN feature}, while the text feature is the same.}
	\begin{center}
		\scalebox{0.9}{
			\begin{tabular}{|c|c|c|c|c|c|c|} 
				\hline
				\multirow{2}{*}{Method} & \multicolumn{3}{c|}{MAP with hand-crafted feature} & \multicolumn{3}{c|}{MAP with CNN feature}\\
				\cline{2-7}
				& Image$\rightarrow$Text & Text$\rightarrow$Image & Average & Image$\rightarrow$Text & Text$\rightarrow$Image & Average \\
				\hline
				\textbf{Our CCL Approach} & \textbf{0.359} & \textbf{0.346} & \textbf{0.353} & \textbf{0.566} & \textbf{0.560} & \textbf{0.563} \\
				CMDN \cite{DBLP:conf/ijcai/PengHQ16} & 0.334 & 0.333 & 0.334 &0.534	 &0.534 &0.534	 \\
				%LGCFL \cite{DBLP:journals/tmm/KangXLXP15} & 0.321 & 0.311 & 0.316 &0.323 &0.326 &0.325\\
				LGCFL \cite{DBLP:journals/tmm/KangXLXP15} & 0.328 & 0.312 & 0.320 &0.539 &0.525 &0.532\\
				JRL \cite{ZhaiTCSVT2014JRL} & 0.300 & 0.286& 0.293 &0.504 &0.489 &0.497\\
				DCCA \cite{DBLP:conf/cvpr/YanM15} & 0.252 & 0.247& 0.250 &0.456 &0.462 &0.459\\
				Corr-AE \cite{feng12014cross} & 0.268 & 0.273& 0.271 &0.489 &0.484 &0.487\\
				Multimodal DBN \cite{srivastava2012learning} & 0.197 & 0.183& 0.190 &0.477 &0.424 &0.451\\
				Bimodal AE \cite{ngiam32011multimodal} & 0.245 & 0.256& 0.251 &0.456 &0.470 &0.463\\
				GMM+HGLMM \cite{DBLP:conf/cvpr/KleinLSW15} & 0.282 & 0.271& 0.277 &0.536 &0.519 &0.528\\
				Multi-label CCA \cite{DBLP:conf/iccv/RanjanRJ15} & 0.262 & 0.257& 0.260 &0.433 &0.434 &0.434\\
				MACC \cite{DBLP:conf/cvpr/TranBC16} & 0.360 & 0.344& 0.352 &0.559 &0.530 &0.545\\
				KCCA(Gaussian) \cite{DBLP:journals/neco/HardoonSS04} & 0.233 & 0.249& 0.241 &0.361 &0.325 &0.343\\
				KCCA(Poly) \cite{DBLP:journals/neco/HardoonSS04} & 0.207 & 0.191& 0.199 &0.209 &0.192 &0.201\\
				CFA \cite{LiMM03CFA} & 0.187 & 0.216& 0.202 &0.351 &0.340 &0.346\\
				CCA \cite{HotelingBiometrika36RelationBetweenTwoVariates} & 0.105 & 0.104& 0.105 &0.169 &0.151 &0.160 \\
				\hline
				
			\end{tabular} 
		}
	\end{center}
	\vspace{-0.5cm}
	\label{table:pascal}
\end{table*}

\begin{table*}[htb]
	\caption{The MAP scores of \textbf{All-modal Retrieval} for our CCL approach and the compared methods on \textbf{Pascal Sentence} dataset. We test two types of image feature: \textbf{Hand-crafted feature} and \textbf{CNN feature}, while the text feature is the same.}
	\begin{center}
		\scalebox{0.9}{
			\begin{tabular}{|c|c|c|c|c|c|c|} 
				\hline
				\multirow{2}{*}{Method} & \multicolumn{3}{c|}{MAP with hand-crafted feature} & \multicolumn{3}{c|}{MAP with CNN feature}\\
				\cline{2-7}
				& Image$\rightarrow$All & Text$\rightarrow$All & Average & Image$\rightarrow$All & Text$\rightarrow$All & Average \\
				\hline
				\textbf{Our CCL Approach} & \textbf{0.352} & \textbf{0.516} & \textbf{0.434} & \textbf{0.554} & \textbf{0.615} & \textbf{0.585} \\
				CMDN \cite{DBLP:conf/ijcai/PengHQ16} & 0.328 & 0.497 & 0.413 &0.532	 &0.604 &0.568 \\
				LGCFL \cite{DBLP:journals/tmm/KangXLXP15} & 0.314 & 0.505 & 0.410 &0.530 &0.604 &0.567\\
				JRL \cite{ZhaiTCSVT2014JRL} & 0.316 & 0.459& 0.388 &0.501 &0.563 &0.532\\
				DCCA \cite{DBLP:conf/cvpr/YanM15} & 0.276 & 0.378& 0.327 &0.465 &0.556 &0.511\\
				Corr-AE \cite{feng12014cross} & 0.305 & 0.367& 0.336 &0.475 &0.558 &0.517\\
				Multimodal DBN \cite{srivastava2012learning} & 0.208 & 0.323& 0.266 &0.459 &0.413 &0.436\\
				Bimodal AE \cite{ngiam32011multimodal} & 0.263 & 0.417& 0.340 &0.466 &0.558 &0.512\\
				GMM+HGLMM \cite{DBLP:conf/cvpr/KleinLSW15} & 0.304 & 0.383& 0.344 &0.513 &0.597 &0.555\\
				Multi-label CCA \cite{DBLP:conf/iccv/RanjanRJ15} & 0.304 & 0.480& 0.392 &0.459 &0.527 &0.493\\
				MACC \cite{DBLP:conf/cvpr/TranBC16} & 0.348 & 0.314& 0.331 &0.538 &0.497 &0.518\\
				KCCA(Gaussian) \cite{DBLP:journals/neco/HardoonSS04} & 0.224 & 0.416& 0.320 &0.423 &0.540 &0.482\\
				KCCA(Poly) \cite{DBLP:journals/neco/HardoonSS04} & 0.218 & 0.446& 0.332 &0.335 &0.261 &0.298\\
				CFA \cite{LiMM03CFA} & 0.206 & 0.395& 0.301 &0.384 &0.427 &0.406\\
				CCA \cite{HotelingBiometrika36RelationBetweenTwoVariates} & 0.196 & 0.226& 0.211 &0.334 &0.232 &0.283 \\
				\hline
				
			\end{tabular} 
		}
	\end{center}
	%\vspace{-0.5cm}
	\label{table:pascal_all}
\end{table*}

\begin{table*}[htb]
	\caption{The MAP scores of \textbf{Bi-modal Retrieval} for our CCL approach and the compared methods on \textbf{NUS-WIDE} dataset. We test two types of image feature: \textbf{Hand-crafted feature} and \textbf{CNN feature}, while the text feature is the same.}
	\begin{center}
		\scalebox{0.9}{
			\begin{tabular}{|c|c|c|c|c|c|c|} 
				\hline
				\multirow{2}{*}{Method} & \multicolumn{3}{c|}{MAP with hand-crafted feature} & \multicolumn{3}{c|}{MAP with CNN feature}\\
				\cline{2-7}
				& Image$\rightarrow$Text & Text$\rightarrow$Image & Average & Image$\rightarrow$Text & Text$\rightarrow$Image & Average \\
				\hline
				\textbf{Our CCL Approach} & \textbf{0.513} & \textbf{0.489} & \textbf{0.501} & \textbf{0.671} & \textbf{0.676} & \textbf{0.674} \\
				CMDN \cite{DBLP:conf/ijcai/PengHQ16} & 0.471 & 0.479 & 0.475 &0.643 &0.626 &0.635 \\
				LGCFL \cite{DBLP:journals/tmm/KangXLXP15} & 0.411 & 0.304 & 0.358 &0.512 &0.600 &0.556\\
				JRL \cite{ZhaiTCSVT2014JRL} & 0.468 & 0.442& 0.455 &0.615 &0.592 &0.604\\
				DCCA \cite{DBLP:conf/cvpr/YanM15} & 0.285 & 0.263& 0.274 &0.475 &0.500 &0.488\\
				Corr-AE \cite{feng12014cross} & 0.308 & 0.336& 0.322 &0.391 &0.429 &0.410\\
				Multimodal DBN \cite{srivastava2012learning} & 0.226 & 0.275& 0.251 &0.213 &0.336 &0.274\\
				Bimodal AE \cite{ngiam32011multimodal} & 0.250 & 0.285& 0.268 &0.307 &0.396 &0.352\\
				GMM+HGLMM \cite{DBLP:conf/cvpr/KleinLSW15} & 0.358 & 0.382& 0.370 &0.588 &0.565 &0.576\\
				Multi-label CCA \cite{DBLP:conf/iccv/RanjanRJ15} & 0.330 & 0.307& 0.319 &0.447 &0.481 &0.464\\
				MACC \cite{DBLP:conf/cvpr/TranBC16} & 0.248 & 0.257& 0.253 &0.492 &0.498 &0.495\\
				KCCA(Gaussian) \cite{DBLP:journals/neco/HardoonSS04} & 0.303 & 0.267& 0.285 &0.348 &0.481 &0.415\\
				KCCA(Poly) \cite{DBLP:journals/neco/HardoonSS04} & 0.221 & 0.215& 0.218 &0.226 &0.243 &0.235\\
				CFA \cite{LiMM03CFA} & 0.290 & 0.283& 0.287 &0.358 &0.361 &0.360\\
				CCA \cite{HotelingBiometrika36RelationBetweenTwoVariates} & 0.233 & 0.229& 0.231 &0.244 &0.275 &0.260 \\
				\hline
				
			\end{tabular} 
		}
	\end{center}
	\vspace{-0.5cm}
	\label{table:nusL}
\end{table*}

\begin{table*}[htb]
	\caption{The MAP scores of \textbf{All-modal Retrieval} for our CCL approach and the compared methods on \textbf{NUS-WIDE} dataset. We test two types of image feature: \textbf{Hand-crafted feature} and \textbf{CNN feature}, while the text feature is the same.}
	\begin{center}
		\scalebox{0.9}{
			\begin{tabular}{|c|c|c|c|c|c|c|} 
				\hline
				\multirow{2}{*}{Method} & \multicolumn{3}{c|}{MAP with hand-crafted feature} & \multicolumn{3}{c|}{MAP with CNN feature}\\
				\cline{2-7}
				& Image$\rightarrow$All & Text$\rightarrow$All & Average & Image$\rightarrow$All & Text$\rightarrow$All & Average \\
				\hline
				\textbf{Our CCL Approach} & \textbf{0.456} & \textbf{0.551} & \textbf{0.504} & \textbf{0.684} & \textbf{0.646} & \textbf{0.665} \\
				CMDN \cite{DBLP:conf/ijcai/PengHQ16} & 0.436 & 0.371 & 0.404 &0.542 &0.579 &0.561 \\
				LGCFL \cite{DBLP:journals/tmm/KangXLXP15} & 0.291 & 0.476 & 0.384 &0.542 &0.588 &0.565\\
				JRL \cite{ZhaiTCSVT2014JRL} & 0.423 & 0.502& 0.463 &0.615 &0.502 &0.559\\
				DCCA \cite{DBLP:conf/cvpr/YanM15} & 0.261 & 0.300& 0.281 &0.508 &0.464 &0.486\\
				Corr-AE \cite{feng12014cross} & 0.272 & 0.305& 0.289 &0.399 &0.381 &0.390\\
				Multimodal DBN \cite{srivastava2012learning} & 0.234 & 0.245& 0.240 &0.218 &0.268 &0.243\\
				Bimodal AE \cite{ngiam32011multimodal} & 0.247 & 0.280& 0.264 &0.319 &0.291 &0.305\\
				GMM+HGLMM \cite{DBLP:conf/cvpr/KleinLSW15} & 0.393 & 0.524& 0.459 &0.584 &0.550 &0.565\\
				Multi-label CCA \cite{DBLP:conf/iccv/RanjanRJ15} & 0.290 & 0.382& 0.336 &0.440 &0.452 &0.446\\
				MACC \cite{DBLP:conf/cvpr/TranBC16} & 0.248 & 0.252& 0.250 &0.450 &0.504 &0.478\\
				KCCA(Gaussian) \cite{DBLP:journals/neco/HardoonSS04} & 0.248 & 0.308& 0.278 &0.318 &0.363 &0.341\\
				KCCA(Poly) \cite{DBLP:journals/neco/HardoonSS04} & 0.228 & 0.229& 0.229 &0.291 &0.240 &0.266\\
				CFA \cite{LiMM03CFA} & 0.251 & 0.281& 0.266 &0.375 &0.320 &0.348\\
				CCA \cite{HotelingBiometrika36RelationBetweenTwoVariates} & 0.232 & 0.253& 0.243 &0.245 &0.269 &0.257 \\
				\hline
				
			\end{tabular} 
		}
	\end{center}
	\vspace{-0.5cm}
	\label{table:nusL_all}
\end{table*}

\begin{table}[htb]
	\caption{The Recall scores of \textbf{Image Annotation} for our CCL approach and the compared methods on \textbf{Flickr-30K} dataset. We test two types of image feature: \textbf{Hand-crafted feature} and \textbf{CNN feature}, while the text feature is the same.}
	\begin{center}
		\scalebox{0.85}{
			\begin{tabular}{|c|c|c|c|c|c|c|} 
				\hline
				\multirow{2}{*}{Method} & \multicolumn{3}{c|}{Hand-crafted feature} & \multicolumn{3}{c|}{CNN feature}\\
				\cline{2-7}
				& R@1 & R@5 & R@10 & R@1 & R@5 & R@10 \\
				\hline
				\textbf{Our CCL Approach} & \textbf{0.088} & \textbf{0.276} & \textbf{0.393} & \textbf{0.377} & \textbf{0.694} & \textbf{0.811} \\
				DCCA \cite{DBLP:conf/cvpr/YanM15} & 0.040 & 0.171& 0.265 &0.279 &0.569 &0.682\\
				Corr-AE \cite{feng12014cross} & 0.054 & 0.195& 0.298 &0.303 &0.615 &0.740\\
				Multimodal DBN \cite{srivastava2012learning} & 0.009 & 0.046& 0.088 &0.064 &0.194 &0.296\\
				Bimodal AE \cite{ngiam32011multimodal} & 0.045 & 0.136& 0.199 &0.127 &0.324 &0.452\\
				GMM+HGLMM \cite{DBLP:conf/cvpr/KleinLSW15} & 0.031 & 0.113& 0.160 &0.350 &0.620 &0.738\\
				%MACC \cite{DBLP:conf/cvpr/TranBC16} & 0.058 & 0.188& 0.305 &0.360 &0.665 &0.785\\
				MACC \cite{DBLP:conf/cvpr/TranBC16} & 0.058 & 0.188& 0.305 &0.139 &0.341 &0.463\\
				KCCA(Gaussian) \cite{DBLP:journals/neco/HardoonSS04} & 0.003 & 0.009& 0.015 &0.108 &0.281 &0.399\\
				KCCA(Poly) \cite{DBLP:journals/neco/HardoonSS04} & 0.002 & 0.006& 0.013 &0.006 &0.036 &0.066\\
				CFA \cite{LiMM03CFA} & 0.055 & 0.159& 0.252 &0.192 &0.449 &0.574\\
				CCA \cite{HotelingBiometrika36RelationBetweenTwoVariates} & 0.006 & 0.052& 0.082 &0.076 &0.205 &0.302 \\
				\hline
				
			\end{tabular} 
		}
	\end{center}
	\vspace{-0.5cm}
	\label{table:flickr_IA}
\end{table}

\begin{table}[htb]
	\caption{The Recall scores of \textbf{Image Retrieval} for our CCL approach and the compared methods on \textbf{Flickr-30K} dataset. We test two types of image feature: \textbf{Hand-crafted feature} and \textbf{CNN feature}, while the text feature is the same.}
	\begin{center}
		\scalebox{0.85}{
			\begin{tabular}{|c|c|c|c|c|c|c|} 
				\hline
				\multirow{2}{*}{Method} & \multicolumn{3}{c|}{Hand-crafted feature} & \multicolumn{3}{c|}{CNN feature}\\
				\cline{2-7}
				& R@1 & R@5 & R@10 & R@1 & R@5 & R@10  \\
				\hline
				\textbf{Our CCL Approach} & \textbf{0.090} & \textbf{0.253} & \textbf{0.361} & \textbf{0.373} & \textbf{0.684} & \textbf{0.800} \\
				DCCA \cite{DBLP:conf/cvpr/YanM15} & 0.053 & 0.157& 0.274 &0.268 &0.529 &0.669\\
				Corr-AE \cite{feng12014cross} & 0.058 & 0.180& 0.294 &0.238 &0.575 &0.707\\
				Multimodal DBN \cite{srivastava2012learning} & 0.026 & 0.094& 0.152 &0.047 &0.151 &0.232\\
				Bimodal AE \cite{ngiam32011multimodal} & 0.029 & 0.103& 0.173 &0.110 &0.328 &0.450\\
				GMM+HGLMM \cite{DBLP:conf/cvpr/KleinLSW15} & 0.031 & 0.105& 0.158 &0.250 &0.527 &0.660\\
				MACC \cite{DBLP:conf/cvpr/TranBC16} & 0.056 & 0.186& 0.293 &0.353 &0.660 &0.782\\
				KCCA(Gaussian) \cite{DBLP:journals/neco/HardoonSS04} & 0.004 & 0.020& 0.041 &0.158 &0.400 &0.543\\
				KCCA(Poly) \cite{DBLP:journals/neco/HardoonSS04} & 0.003 & 0.007& 0.013 &0.015 &0.055 &0.087\\
				CFA \cite{LiMM03CFA} & 0.067 & 0.199& 0.295 &0.242 &0.566 &0.683\\
				CCA \cite{HotelingBiometrika36RelationBetweenTwoVariates} & 0.011 & 0.065& 0.104 &0.091 &0.268 &0.390 \\
				\hline
				
			\end{tabular} 
		}
	\end{center}
	\vspace{-0.5cm}
	\label{table:flickr_IR}
\end{table}

\begin{table}[htb]
	\caption{The Recall scores of \textbf{Image Annotation} for our CCL approach and the compared methods on \textbf{MS-COCO} dataset. We test two types of image feature: \textbf{Hand-crafted feature} and \textbf{CNN feature}, while the text feature is the same.}
	\begin{center}
		\scalebox{0.85}{
			\begin{tabular}{|c|c|c|c|c|c|c|} 
				\hline
				\multirow{2}{*}{Method} & \multicolumn{3}{c|}{Hand-crafted feature} & \multicolumn{3}{c|}{CNN feature}\\
				\cline{2-7}
				& R@1 & R@5 & R@10 & R@1 & R@5 & R@10 \\
				\hline
				\textbf{Our CCL Approach} & \textbf{0.063} & \textbf{0.196} & \textbf{0.298} & \textbf{0.186} & \textbf{0.474} & \textbf{0.625} \\
				DCCA \cite{DBLP:conf/cvpr/YanM15} & 0.023 & 0.071& 0.116 &0.069 &0.211 &0.318\\
				Corr-AE \cite{feng12014cross} & 0.024 & 0.090& 0.147 &0.154 &0.397 &0.532\\
				Multimodal DBN \cite{srivastava2012learning} & 0.013 & 0.047& 0.082 &0.054 &0.194 &0.292\\
				Bimodal AE \cite{ngiam32011multimodal} & 0.015 & 0.062& 0.104 &0.063 &0.220 &0.347\\
				GMM+HGLMM \cite{DBLP:conf/cvpr/KleinLSW15} & 0.019 & 0.071& 0.115 &0.173 &0.390 &0.502\\
				MACC \cite{DBLP:conf/cvpr/TranBC16} & 0.012 & 0.041& 0.069 &0.056 &0.167 &0.244\\
				KCCA(Gaussian) \cite{DBLP:journals/neco/HardoonSS04} & 0.011 & 0.045& 0.074 &0.072 &0.202 &0.305\\
				KCCA(Poly) \cite{DBLP:journals/neco/HardoonSS04} & 0.004 & 0.013& 0.023 &0.003 &0.015 &0.029\\
				CFA \cite{LiMM03CFA} & 0.021 & 0.071& 0.129 &0.086 &0.258 &0.371\\
				CCA \cite{HotelingBiometrika36RelationBetweenTwoVariates} & 0.003 & 0.014& 0.023 &0.041 &0.142 &0.226 \\
				\hline
				
			\end{tabular} 
		}
	\end{center}
	\vspace{-0.5cm}
	\label{table:coco_IA}
\end{table}

\begin{table}[htb]
	\caption{The Recall scores of \textbf{Image Retrieval} for our CCL approach and the compared methods on \textbf{MS-COCO} dataset. We test two types of image feature: \textbf{Hand-crafted feature} and \textbf{CNN feature}, while the text feature is the same.}
	\begin{center}
		\scalebox{0.85}{
			\begin{tabular}{|c|c|c|c|c|c|c|} 
				\hline
				\multirow{2}{*}{Method} & \multicolumn{3}{c|}{Hand-crafted feature} & \multicolumn{3}{c|}{CNN feature}\\
				\cline{2-7}
				& R@1 & R@5 & R@10 & R@1 & R@5 & R@10  \\
				\hline
				\textbf{Our CCL Approach} & \textbf{0.064} & \textbf{0.197} & \textbf{0.296} & \textbf{0.196} & \textbf{0.469} & \textbf{0.623} \\
				DCCA \cite{DBLP:conf/cvpr/YanM15} & 0.021 & 0.075& 0.118 &0.066 &0.209 &0.322\\
				Corr-AE \cite{feng12014cross} & 0.026 & 0.094& 0.151 &0.138 &0.353 &0.478\\
				Multimodal DBN \cite{srivastava2012learning} & 0.014 & 0.054& 0.090 &0.046 &0.155 &0.240\\
				Bimodal AE \cite{ngiam32011multimodal} & 0.015 & 0.062& 0.099 &0.054 &0.178 &0.283\\
				GMM+HGLMM \cite{DBLP:conf/cvpr/KleinLSW15} & 0.022 & 0.072& 0.115 &0.108 &0.283 &0.401\\
				MACC \cite{DBLP:conf/cvpr/TranBC16} & 0.015 & 0.058& 0.089 &0.155 &0.370 &0.490\\
				KCCA(Gaussian) \cite{DBLP:journals/neco/HardoonSS04} & 0.003 & 0.015& 0.019 &0.020 &0.074 &0.122\\
				KCCA(Poly) \cite{DBLP:journals/neco/HardoonSS04} & 0.005 & 0.017& 0.027 &0.016 &0.061 &0.102\\
				CFA \cite{LiMM03CFA} & 0.045 & 0.133& 0.192 &0.150 &0.381 &0.514\\
				CCA \cite{HotelingBiometrika36RelationBetweenTwoVariates} & 0.004 & 0.022& 0.037 &0.041 &0.155 &0.251 \\
				\hline
				
			\end{tabular} 
		}
	\end{center}
	
	\label{table:coco_IR}
\end{table}

\section{Experiments}

We conduct experiments on 6 widely-used cross-modal datasets compared with 13 state-of-the-art methods to verify its effectiveness.
%In this paper, we only conduct experiments with image and text, because of the limitation that the cross-modal datasets we used only contain these two modalities. However, our approach can take other modalities such as audio and video as input, and perform the cross-modal retrieval between them. 
%And we compare our proposed CCL approach with 13 state-of-the-art methods to verify its effectiveness. 
In addition, we also present comprehensive experimental analysis, including network and parameter analysis, execution time and baseline experiments to verify separate contribution of each component in our approach.
%In addition, we also conduct two groups of baseline experiments to further verify separate contribution of each part in our approach: (1) Cross-modal retrieval with only coarse-grained original instances or only fine-grained patches, (2) Cross-modal retrieval with only inter-modality correlation or only intra-modality correlation in the first learning stage.

%\begin{figure}[!t]
%	\centering
%	\includegraphics[width=3.2in]{Figure/WikiPedia_Ex_v5-eps-converted-to.pdf}
%	\caption{Examples of 2 categories from the Wikipedia dataset, which are Biology and Sport.}
%	\label{fig_wiki_ex}
%\end{figure}

\subsection{Datasets}

%Here we briefly introduce 6 widely-used cross-modal datasets adopted in the experiment. For fair and objective comparison purpose, we exactly follow the dataset partition strategy of \cite{feng12014cross,DBLP:conf/ijcai/PengHQ16} in the experiment to divide the dataset into 3 subsets: Training set, testing set and validation set. 

\textbf{Wikipedia dataset} \cite{RasiwasiaMM10SemanticCCA} is the most widely-used dataset for cross-modal retrieval. This dataset consists of 2,866 image/text pairs of 10 categories, and is randomly split following \cite{feng12014cross,DBLP:conf/ijcai/PengHQ16}: 2,173 pairs for training, 231 pairs for validation and 462 pairs for testing.% Some examples are shown in Figure \ref{fig_wiki_ex}. 

\textbf{NUS-WIDE dataset} \cite{NUSWIDE} is a web image dataset for media search, which consists of about 270,000 images with their tags categorized into 81 classes. 
%Because of overlapping among the categories, o
Only the images exclusively belonging to one of the 10 largest categories in NUS-WIDE dataset are selected for experiments following \cite{DBLP:conf/aaai/ZhuangWWZL13}, and each image along with its corresponding tags is viewed together as an image/text pair with unique class label. Finally, there are about 70,000 image/text pairs, where 
%Each image along with its corresponding tags is viewed together as an image/text pair. Because of overlapping among the categories, 10 largest categories are selected following \cite{DBLP:conf/aaai/ZhuangWWZL13}, which include about 70,000 image/text pairs with unique class label. 
training set consists of 42,941 pairs, testing set is with 23,661 pairs, while 5,000 pairs are in validation set.

\textbf{NUS-WIDE-10K dataset} \cite{NUSWIDE} has totally 10,000 image/text pairs selected evenly from the 10 largest categories of NUS-WIDE dataset, which are animal, cloud and so on. The dataset is split into three subsets following \cite{feng12014cross,DBLP:conf/ijcai/PengHQ16}: Training set with 8,000 pairs, testing set with 1,000 pairs and validation set with 1,000 pairs. 

\textbf{Pascal Sentence dataset} \cite{rashtchian2010collecting} is generated from 2008 PASCAL development kit. This dataset contains 1,000 images which are evenly categorized into 20 categories, and each image has 5 corresponding sentences which make up one document. For each category, 40 documents are selected for training, 5 documents for testing and 5 documents for validation following \cite{feng12014cross,DBLP:conf/ijcai/PengHQ16}. 

\textbf{Flickr-30K dataset} \cite{DBLP:journals/tacl/YoungLHH14} consists of 31,784 images from the Flickr.com website. Each image is annotated by 5 sentences generated by the crowdsourcing service with different annotators. Following \cite{DBLP:conf/cvpr/TranBC16}, there are 1,000 pairs in testing set and 1,000 pairs in validation set, while the rest are training set.

\textbf{MS-COCO dataset} \cite{DBLP:conf/eccv/LinMBHPRDZ14} contains 123,287 images and their annotated sentences. The annotations of images are also generated by crowdsourcing via Amazon Mechanical Turk, and each image is annotated by 5 independent sentences provided by 5 users. Following \cite{DBLP:conf/cvpr/KleinLSW15}, there are both 5,000 pairs split randomly as testing set and validation set, while the rest are training set.

\subsection{Patch Segmentation}
%For fully taking advantage of the intrinsic semantic correlation and rich complementary information provided by the instance patches, we divide the image and text into several patches. 
For the image, all 6 datasets share the same segmentation method. Selective search \cite{DBLP:journals/ijcv/UijlingsSGS13} is adopted to generate thousands of region proposals for one image. We also consider the criterion of non-overlapping when selecting the image patches. Specifically, we set the threshold of Intersect over Union (IoU) between different patches as 0.7. When the IoU of two patches is larger than 0.7, we discard the smaller one. Finally, the 10 largest patches are automatically selected, which have higher probability to contain visual objects with rich fine-grained information.

%and then up to largest 10 patches are automatically selected. 
%For the text instances, the segmentation methods for different datasets are different. 
Besides, the form of text varies among different datasets, so different segmentation methods are adopted. 
%For example, the texts in Wikipedia dataset are articles, while the texts in NUSWIDE-10k dataset are made up by tags. In this case, different text segmentation methods are adopted for different datasets. Specifically, 
Specifically, for Wikipedia dataset, each text instance is in the form of an article consisting of several paragraphs. So we split each text instance by the paragraph where each paragraph contains relevant content. For Pascal Sentence, Flickr-30K and MS-COCO datasets, each text instance consists of 5 sentences. Therefore, each sentence that describes the corresponding image is treated as a patch. As for NUS-WIDE dataset and its subset NUS-WIDE-10K dataset, the text instances are independent tags, rather than sentences or paragraphs. We first arrange the tags with alphabetical order. 
Then, we analyze the distribution over the number of tags associated with each image in the dataset, which shows that more than half of images in the dataset have more than 4 tags. Thus, we intuitively divide each text instance into 4 patches for generality.
%Then we divide them into 4 patches for uniformity where each patch has the same number of words, which can make generated patches with more generality. 
For the images with less than 4 tags, each annotated tag is regarded as one patch. So in this case, the number of patches is less than 4 for a text instance.  
%The texts of Wikipedia dataset are in the form of articles with several paragraphs, thus we divide them by paragraph. The texts in Pascal Sentence are made up by several sentences, so it is divided by each sentence. Since the text instances in NUS-WIDE-10k dataset are made up of several tags which has no context relationship, we divide them by word if the number of words is less than 4, otherwise divide them into 4 patches for uniformity where each patch has the same number of words. 

\subsection{Feature extraction}

We extract the hand-crafted features exactly the same as \cite{feng12014cross} on both the original instances and their patches for fair comparison. 
For Wikipedia dataset, the image features are the concatenation of three parts with totally 2,296 dimensions: 1,000 dimensional Pyramid Histogram of Words, 512 dimensional GIST, and 784 dimensional MPEG-7 descriptor. The text feature is 3,000 dimensional bag-of-words vector. 
For NUS-WIDE and NUS-WIDE-10K datasets, the image feature has totally 1,134 dimensions, which is the concatenation of 64 dimensional color histogram, 144 dimensional color correlogram, 73 dimensional edge direction histogram, 128 dimensional wavelet texture, 225 dimensional block-wise color moments and 500 dimensional SIFT based bag-of-words features. The text feature is 1,000 dimensional bag-of-words vector.
For Pascal Sentence, Flickr-30K and MS-COCO datasets, the image feature is the same as Wikipedia dataset with the totally 2,296 dimensions. And the text feature of Pascal Sentences dataset is 1,000 dimensional bag-of-words vector, while for Flickr-30K and MS-COCO datasets, it is 3,000 dimensions.
In addition, CNN feature has shown its effectiveness for image representation, so we also use CNN image feature in the experiments. Specifically, the adopted CNN feature has 4,096 dimensions extracted by the fc7 layer of VGGNet \cite{simon2016cnnmodels} for all compared methods on 6 datasets.% for further verifying the effectiveness of our approach.

It is noted that for each dataset, %the feature extraction on the patches is same as that on the original instances.
we extract the same type of feature with the same number of dimensions for both original instances and patches. Taking image for example, we first generate $l$ patches $\{p_1, p_2, ... , p_l\}$ for an original image $I$, and then the features of these $l$ patches and $I$ are extracted separately in the same way.

\subsection{Details of the Network}

%We introduce the details of the network here. 
Our CCL approach has two learning stages as shown in Figure \ref{fig:network}. In the first learning stage, four two-layer DBN are used to model the image features and text features of original instances and their patches separately. The DBN for image has 2,048 dimensions in the first layer and 1,024 dimensions in the second layer, and the DBN for texts has 1,024 dimensions in both layers. After the two-layer DBN, representations of patches which belong to the same instance are averagely fused to one vector. 
Then for intra-modality and inter-modality correlation modeling on original instances, the two-pathway network has three layers with 1,024 dimensions linked at the top code layer for joint optimization, which is the same for patches. 
Finally, we use the joint RBM to fuse the representations of original instance and patches for image and text respectively. The output dimension of joint RBM is 2,048. On the top of joint RBM, a three-layer feed-forward network is used for further optimization with softmax loss, and the dimensional number of each layer is 1,024. The above networks are implemented with deepnet\footnote{https://github.com/nitishsrivastava/deepnet}.

In the second learning stage, the multi-task learning strategy is adopted where the network is a three-layer fully-connected network implemented by Caffe \cite{jia2014caffe}. 
%For Wikipedia dataset and NUS-WIDE-10k dataset, the network we choose is a three-layer fully-connected network which is implemented by Caffe \cite{jia2014caffe}. For Pascal Sentence dataset, we decrease the number of layers to two to avoid over-fitting. 
All the dimensional numbers of three fully-connected layers are 1,024. The network adopts different loss functions for intra-modality and inter-modality learning tasks, and the fully-connected layer on inter-modality loss branch has 1,024 dimensions, while that on intra-modality loss branch has also 1,024 dimensions. 
%Besides, the range of $\alpha$ in equ. (\ref{equ:ls}) is from 0.5 to 2.5 according to the distance range between image and text features, and it is set to be 1 in our experiments.

The parameters mentioned above are for the Wikipedia dataset. For the fact that the dimensions of input features in different datasets are not the same, the dimensional number in the first layer of DBN is adjusted according to different input dimensions, while the other parameters, such as the dimensional number of each layer, remain the same.
%Specifically, for the hand-crafted image feature of NUS-WIDE dataset, we set the first layer of DBN for images with 1,024 hidden units. And for the CNN feature in different datasets, the first layer of DBN for images has 2,048 hidden units.

Besides, to address the over-fitting issue, we take several measures in the training process as follows: First, dropout layers are inserted appropriately in the proposed network, which can effectively prevent over-fitting problem as indicated in \cite{DBLP:journals/corr/abs-1207-0580}. Second, all the training data is shuffled, while the image data is also augmented with mirrored versions to expand the training set following \cite{DBLP:conf/cvpr/YanM15}. Third, a good initialization for the parameters in the network also takes effect. For example, the weights of the fully-connected layers are initialized to identity matrices and biases to zeros. This aims to make the optimal parameters search from a safe point. 

\subsection{Compared Methods and Evaluation Metric}

We conduct two cross-modal retrieval tasks on Wikipedia, NUS-WIDE, NUS-WIDE-10K and Pascal Sentence datasets following \cite{feng12014cross,DBLP:conf/ijcai/PengHQ16}:
\begin{itemize}
	\item {\textbf{Bi-modal retrieval}.} Retrieving one modality in testing set using a query of another modality, namely retrieving text by image (image$\to$text), and retrieving image by text (text$\to$image).
	\item {\textbf{All-modal retrieval}.} Retrieving all modalities in testing set using a query of any modality, namely retrieving image and text together by an image query (image$\to$all) and a text query (text$\to$all).
\end{itemize}
As for Flickr-30K and MS-COCO datasets, two bi-modal retrieval tasks are conducted following \cite{DBLP:conf/cvpr/YanM15,DBLP:conf/cvpr/KleinLSW15}:
\begin{itemize}
	\item {\textbf{Image annotation}.} Retrieving the groundtruth sentences given a query image (image$\to$text).
	\item {\textbf{Image retrieval}.} Retrieving the groundtruth images given a query text (text$\to$image).
\end{itemize}
%It should be noted that data of different modalities can be converted to common representations with the same number of dimensions in the testing stage. So we can directly compute the similarity between the query and any data (no matter it is image or text) by distance measurement.
%We conduct four retrieval tasks on each dataset mentioned above, namely retrieving text by image (image$\to$text), retrieving image by text (text$\to$image), retrieval among all the modality types by using an image query (image$\to$all) or a text query (text$\to$all). 
We compare our CCL approach with 13 state-of-the-art methods to verify its effectiveness. 
It should be noted that Flickr-30K and MS-COCO datasets have no label annotations available, so some compared methods cannot be conducted including Multi-label CCA, JRL and CMDN, because they all need category information for common representation learning. 
This also leads to some changes on the multi-task framework of our proposed CCL approach. For inter-modality pairwise similarity constraints, although these two datasets have no labels available, there still exist co-existence relationship between image and text because each image has its corresponding text description, which forms an image/text pair. So we construct the similarity matrix $E$ as Eq.(\ref{eq_e}) according to the pairwise co-existence. As for the intra-modality semantic category constraints which need labels, we have to drop this part for comparison on these two datasets.
%CCL also adopts the intra-modality semantic category constraints in the proposed multi-task learning network, which needs label annotation. Therefore, we have to drop the above part for comparison on these two datasets.
In the experiments, our proposed CCL approach and all the compared methods are evaluated by the following steps for fair comparison. (1) Learn projections or deep models for common representation with data in the training set; (2) Convert all data in testing set by learned projections or deep models to common representation with the same number of dimension; 
%Specifically, in CCL the outputs of last fully-connected layer in the multi-task learning network are obtained as common representations; 
(3) Perform retrieval with common representation directly by the same cosine distance. 
For the fact that some compared methods, such as MACC \cite{DBLP:conf/cvpr/TranBC16}, DCCA \cite{DBLP:conf/cvpr/YanM15}, and GMM+HGLMM \cite{DBLP:conf/cvpr/KleinLSW15}, do not report the results on some of our selected datasets, we directly adopt the source codes provided by their authors to obtain the final retrieval results for comparison. For example, the paper of MACC \cite{DBLP:conf/cvpr/TranBC16} only reports the result of image retrieval task with CNN feature on Flickr-30K dataset \cite{DBLP:journals/tacl/YoungLHH14} (see Table \ref{table:flickr_IR}), so we also get the image annotation results and the results of two retrieval tasks on other datasets using the source codes provided by their authors (see Tables \ref{table:wiki} to \ref{table:flickr_IA} and \ref{table:coco_IA} to \ref{table:coco_IR}).

We adopt the mean average precision (MAP) score as the evaluation metric on Wikipedia, NUS-WIDE, NUS-WIDE-10K and Pascal Sentence datasets, which takes the precision and ranking of the returned retrieval results into consideration at the same time. For a set of queries, MAP score is the mean of Average Precision (AP) of each query. 
AP can be calculated by the following formula: 
\begin{gather}
AP= \frac{1}{R}\sum_{k=1}^{n}\frac{R_{k}}{k}\times rel_{k}
\end{gather}
where \emph{n} is the number of retrieval set, \emph{R} means the number of relevant items and \emph{R$_k$} counts the number of relevant items in the top \emph{k} results. When the \emph{k}-th result is relevant, \emph{rel$_k$} is set to be 1, otherwise 0. 
It should be noted that we show the results of MAP score on \emph{all returned results}, which is extensively adopted in cross-modal retrieval task as \cite{RasiwasiaMM10SemanticCCA}, \cite{ZhaiTCSVT2014JRL,DBLP:journals/tmm/KangXLXP15}, and do not adopt only \emph{top 50} like \cite{feng12014cross}. %, instead of MAP for \textit{the top 50 results} adopted in \cite{feng12014cross} for more comprehensive comparison. %but also the MAP@50 which means the MAP scores on the top 50 returned results.
Besides, we also adopt precision-recall and precision-scope curves on large-scale NUS-WIDE dataset for comprehensive evaluation. 

As for Flickr-30K and MS-COCO datasets, because there are no labels available, the MAP score, precision-recall curve and precision-scope curve are not applicable. Instead, we report the score of Recall@K following \cite{DBLP:conf/cvpr/YanM15,DBLP:conf/cvpr/KleinLSW15}, which includes the recall rate at top 1 result (R@1), top 5 results (R@5) and top 10 results (R@10). %It should be noted that these two datasets have no label annotation available, some compared methods cannot be conducted including Multi-label CCA, JRL and CMDN, because they all need category information in the common representation learning. While we also adopt the intra-modality semantic category constraints in the proposed multi-task learning network, which aims to classify the data of each modality into one of $n$ categories by an $n$-way softmax layer. Therefore, we have to drop the appropriate part for comparison on these two datasets.
%Besides, we also adopt precision-recall and precision-scope curves for comprehensive evaluations. Due to the length limitation of this paper, we only show precision-recall and precision-scope curves on NUS-WIDE dataset.

%It should be noted that the original three datasets are all in small-scale, thus only NUS-WIDE dataset with large scale is selected for more comprehensive evaluation for the length limitation of this paper. 
%Also for the fact that Filckr-30K and MS-COCO datasets have different evaluation metric, it is not suitable for precision-scope and precision-recall curves.

\begin{figure*}[!t]
	\centering
		\includegraphics[width=0.9\textwidth]{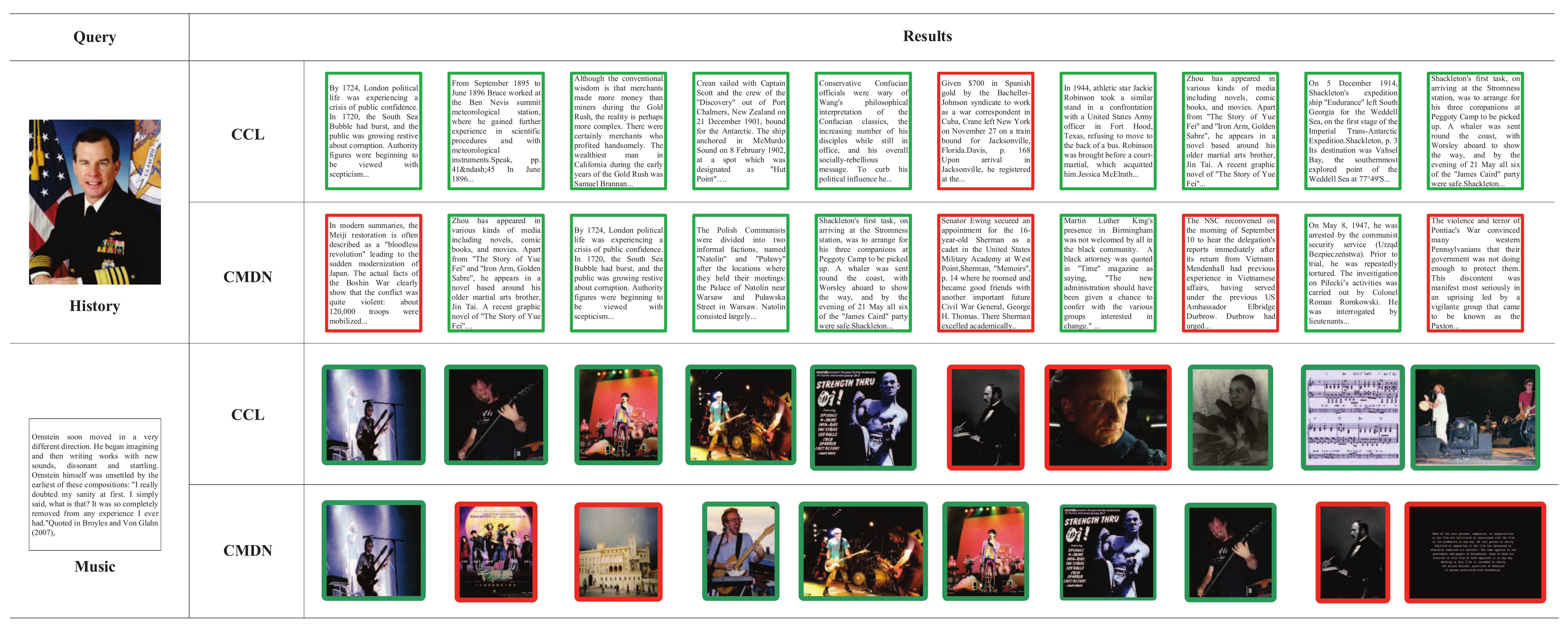}
	\caption{Examples of the bi-modal retrieval results with hand-crafted feature on Wikipedia dataset by our CCL approach and CMDN \cite{DBLP:conf/ijcai/PengHQ16}. It should be noted that, in these examples, the correct retrieval results are with green borders, while the wrong results are with red borders.
	}
	\label{fig_wiki_res}
\end{figure*}

\begin{figure}[!t]
	\centering
	\includegraphics[width=3.35in]{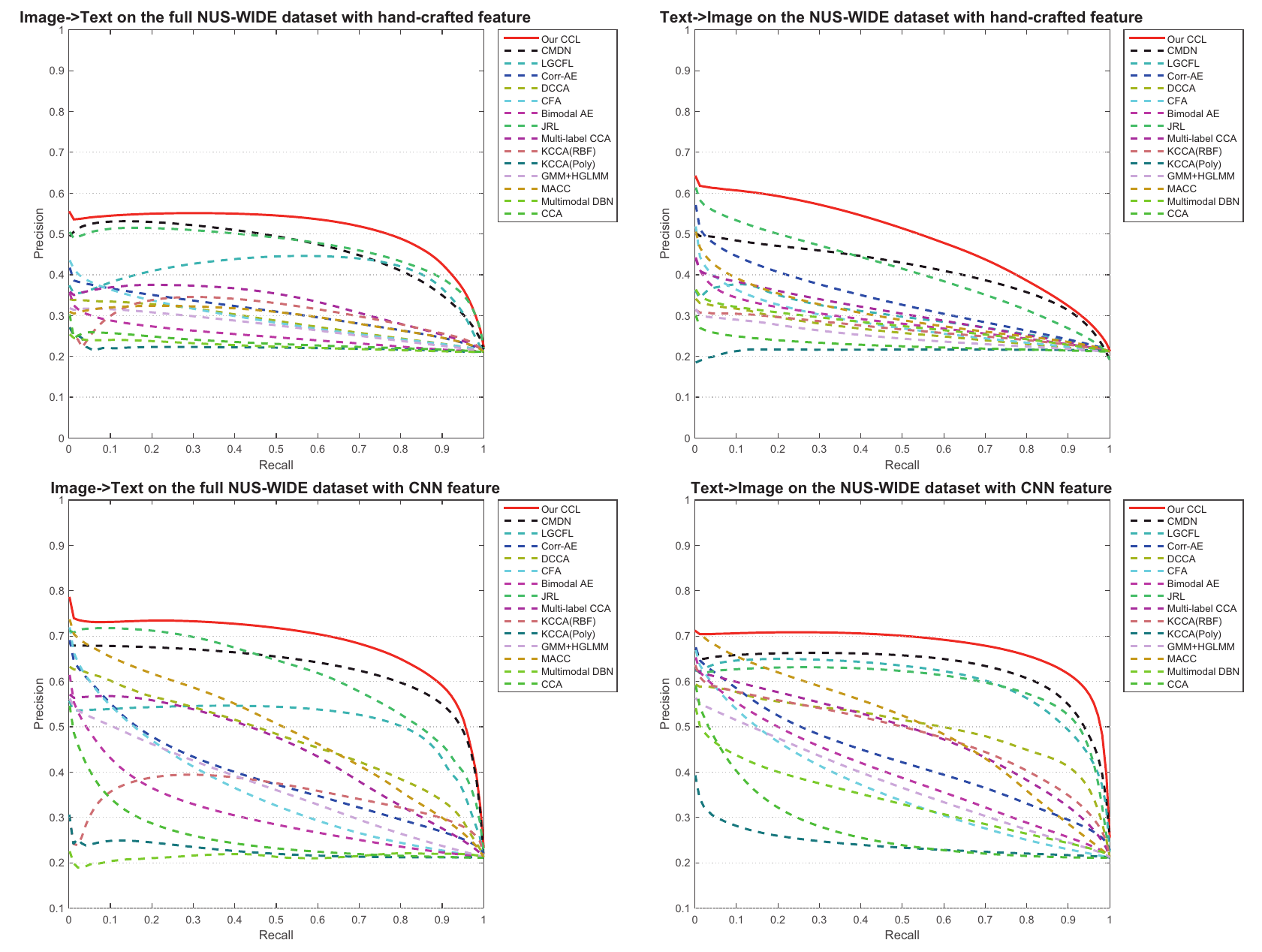}
	\caption{Precision-Recall curves of bi-modal retrieval on NUS-WIDE dataset.}
	\label{fig_pr}
\end{figure}

%\begin{minipage}[c]{\linewidth}
%	\centering
%	\includegraphics[width=0.3\textwidth]{Figure/PS/ps_i2t_1134-eps-converted-to.pdf}
%	\includegraphics[width=0.3\textwidth]{Figure/PS/ps_t2i_1134-eps-converted-to.pdf}\\
%	
%	\includegraphics[width=0.3\textwidth]{Figure/PS/ps_i2t_cnn-eps-converted-to.pdf}
%	\includegraphics[width=0.3\textwidth]{Figure/PS/ps_t2i_cnn-eps-converted-to.pdf}
%%	\caption{Precision-Scope curves on NUS-WIDE dataset.}
%	\label{fig_ps}
%	
%\end{minipage}%

\subsection{Comparison with State-of-the-art Methods}

%\subsubsection{Comparison with Existing Methods}
This part presents the experimental results and analyses on our CCL approach as well as all the compared methods.
As shown in Tables \ref{table:wiki} and \ref{table:wiki_all}, our approach achieves the best results in both bi-modal and all-modal retrieval tasks on Wikipedia dataset with both hand-crafted feature and CNN feature. 
%Compared with our previous work CMDN \cite{DBLP:conf/ijcai/PengHQ16} which gets the best retrieval accuracy in all the compared methods, our CCL approach has improved the average MAP score from 0.359 to 0.389. 
Some retrieval results are shown by our CCL approach and CMDN in Figure \ref{fig_wiki_res}, from which we can see that our proposed CCL approach can effectively reduce the failure cases compared with other methods. And a few categories in this dataset are difficult to be distinguished due to the high-level semantics such as ``history" category, which leads to confusions during retrieval process on both our CCL approach and the compared methods. But our approach still achieves the best retrieval accuracy with the least failure cases.
Besides, we have made classification on the failure cases with the following two types: (1) Failure due to the confusion in the image instance, which indicates that small variance between the image instances of different categories leads to wrong retrieval results; (2) Failure due to the confusion in the text instance, which is similarly caused by small variance between the text instances of different categories.

%Among the compared traditional methods, LGCFL has the best retrieval accuracy, but slightly lower than CMDN based on DNN. The accuracies of four DNN-based methods vary greatly. Multimodel DBN has the worst accuracy among the DNN-based methods, and even lower than most traditional methods except CCA. 

\begin{figure}[!t]
	\centering
	\includegraphics[width=3.3in]{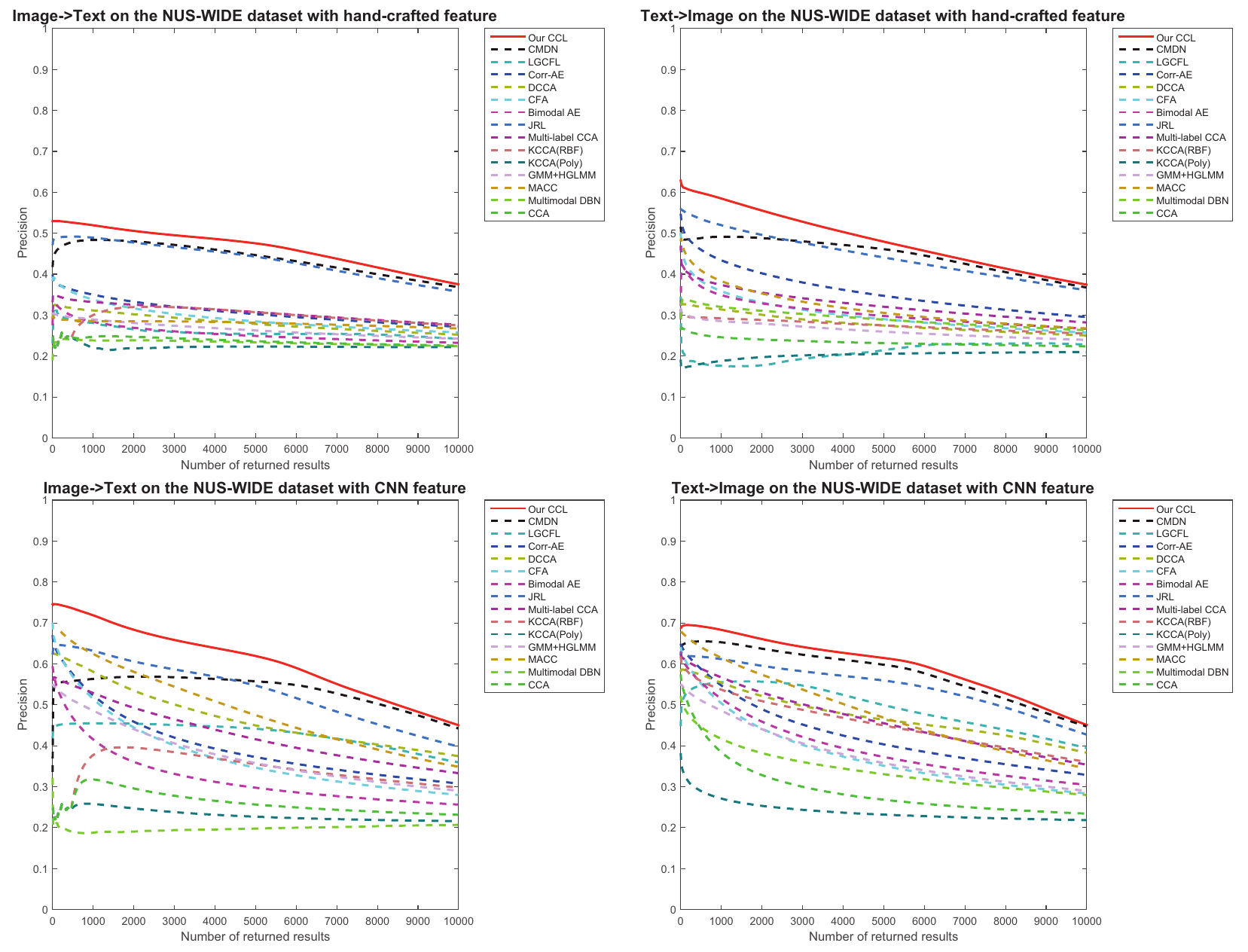}
	\caption{Precision-Scope curves of bi-modal retrieval on NUS-WIDE dataset.}
	\label{fig_ps}
\end{figure}

%As for the all-modal retrieval task shown in Table \ref{table:wiki_all}, our CCL approach still outperforms all the compared method with hand-crafted features.
Besides, the results on other 5 datasets with both hand-crafted feature and CNN feature are shown from Tables \ref{table:nus} to \ref{table:coco_IR}. The trends of the results on these datasets are similar to Wikipedia dataset and our CCL approach keeps the best. 
%Except for the MAP score based on hand-crafted features, we also show the MAP score with CNN image feature on the all datasets. As shown from Tables \ref{table:wiki} to \ref{table:pascal_all}, 
%our CCL approach still outperforms all the compared methods on all 3 datasets. 
%Also we can observe that deep features have greatly boosted the accuracy of both traditional methods and DNN-based methods, comparing with the hand-crafted features.
Also we can observe that the compared methods generally benefit greatly from CNN feature and achieve accuracy improvement. 
%However, there still exist some exceptions. For example, %the performance of bi-modal retrieval between image and text by KCCA (Poly) on Pascal Sentence dataset almost keeps the same with hand-crafted feature, and 
%KCCA (Poly) even shows a clear decrement with CNN feature on the average MAP score of all-modal retrieval on Pascal Sentence dataset. 
Our CCL approach can stably benefit from CNN image feature and keeps the best among all compared methods.
Figures \ref{fig_pr} and \ref{fig_ps} show the precision-recall curves and precision-scope curves of bi-modal retrieval task on NUS-WIDE dataset, which can further verify the effectiveness of our CCL approach.

%In Tables \ref{table:wiki_all}, \ref{table:nus_all} and \ref{table:pascal_all}, we show the MAP score with different types of image features for cross-modal retrieval task, and the queries are set as image or text. Similar with bi-modal retrieval tasks between image and text, we compare our CCL approach with 9 state-of-art methods on all 3 datasets. As we can see, our approach outperforms all the compared methods on all 3 datasets with the two tasks.

As shown from Tables \ref{table:wiki} to \ref{table:coco_IR}, our CCL approach keeps advantage with all 13 compared methods on 6 datasets, and the results show similar trends. 
Among the traditional methods, CFA minimizes the Frobenius norm in transformed domain, KCCA adopts kernel functions, and multi-label CCA considers the high-level semantic information, which makes them perform better than classical baseline CCA. 
It should be noted that KCCA has lower accuracy than CFA in some settings for the fact that KCCA can only learn a coarse association between different modalities. In addition, the performance of different kernel functions may vary greatly. Especially, Poly kernel has lower accuracy than Gaussian kernel on most settings because it cannot effectively handle the large-scale training data.
MACC further embeds the projections into a local context based on KCCA. JRL outperforms the above methods by using semi-supervised regularization and sparse regularization. GMM+HGLMM and LGCFL achieve the best accuracies among the traditional methods in most cases, because GMM+HGLMM combines different kinds of Mixture Models to improve sentence representation, while LGCFL jointly learns basis matrices using a local group based priori.
As for the DNN-based methods, Multimodal DBN has the worst accuracy among them. It is because only a joint distribution is learned on the top of two-layer DBN for each modality, which focuses more on learning the complementarity between different modalities rather than the correlation across them. Bimodal AE and Corr-AE have better accuracies because they jointly consider the reconstruction information. DCCA extends traditional CCA by maximizing the correlation between two separate networks. Our previous work CMDN achieves the best accuracy among the DNN-based methods because of jointly modeling on intra-modality and inter-modality correlation in both two learning stages.
It should be noted that our proposed CCL approach achieves the best accuracy compared with state-of-the-art methods for the following 3 aspects: (1) The multi-grained fusion on both coarse-grained instances and fine-grained patches, while the compared methods only use the original modality instances. (2) The multi-task learning strategy to adaptively balance intra-modality semantic category constraints and inter-modality pairwise similarity constraints, while the compared methods only adopt single-loss regularization failing to effectively exploit and balance the above constraints. (3) Jointly optimizing on the intrinsic intra-modality and inter-modality correlation, while the compared methods ignore the inter-modality correlation in the separate representation learning for the first stage.

\subsection{Experimental Analysis}

\subsubsection{Network and Parameter Analysis}

\begin{figure}[!t]
	\centering
	\includegraphics[width=0.45\textwidth]{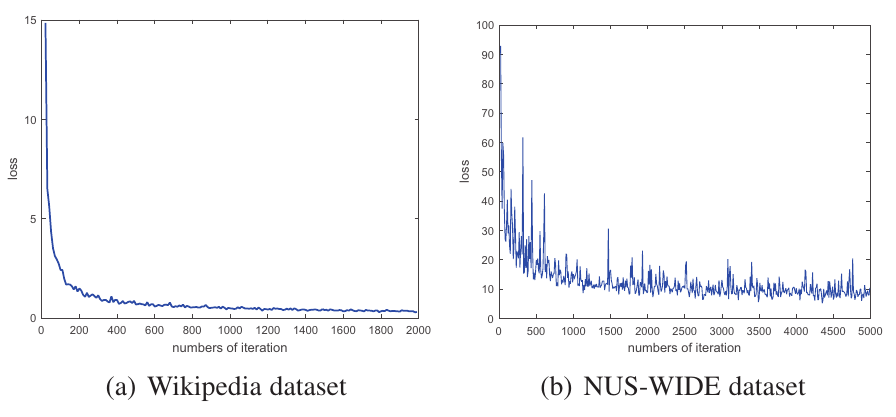}
	\caption{Convergence experiments of our proposed CCL approach conducted on Wikipedia dataset and NUS-WIDE dataset, which show the curves of downtrend on the loss value.
	}
	\label{fig_loss}
\end{figure}

\begin{figure*}[!t]
	\centering
	\includegraphics[width=0.9\textwidth]{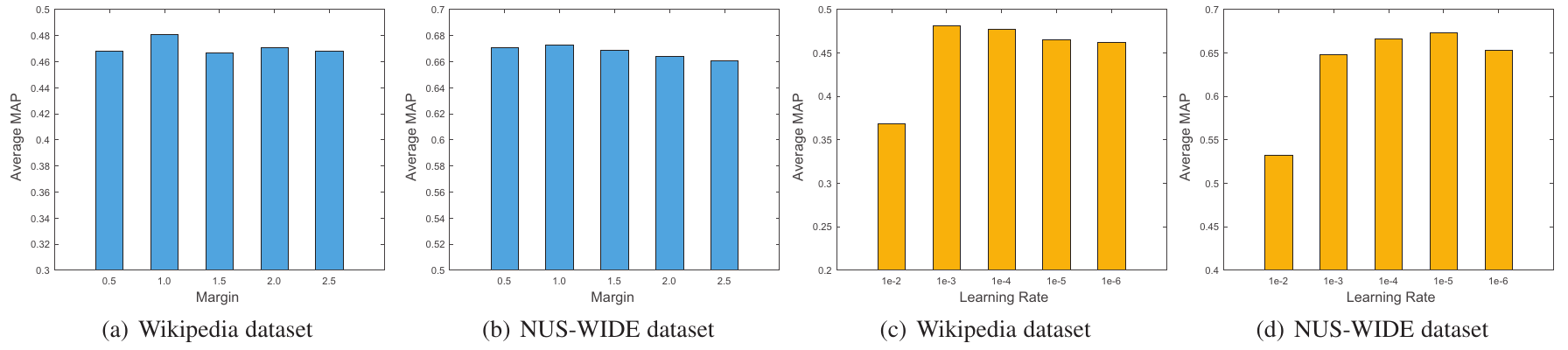}
	\caption{Experiments on the training detail of our proposed CCL approach, including learning rate and the margin parameter $\alpha$, on Wikipedia dataset and NUS-WIDE dataset. It should be noted that we report the average MAP score of Image$\rightarrow$Text and Text$\rightarrow$Image tasks.
	}
	\label{fig_para}
\end{figure*}

For the \textit{\textbf{convergence}} experiments, which are conducted on Wikipedia and NUS-WIDE datasets as examples due to the page limitation, we show the curves of downtrend on the loss value in Figure \ref{fig_loss} to verify the convergence of our proposed approach. We can see that our CCL approach can converge quickly within 2K iterations, which shows its efficiency.
For the \textit{\textbf{training details}} of the network, we conduct the experiments on the effect of some key parameters also on Wikipedia and NUS-WIDE datasets, including learning rate ranging from 1e-2 to 1e-6, and the margin parameter $\alpha$ in the loss function in Eq.(\ref{equ:ls}) ranging from 0.5 to 2.5. The results are shown in Figure \ref{fig_para}. And we can see that our CCL approach gets the best performance at the learning rate of 1e-3 on Wikipedia dataset and 1e-5 on NUS-WIDE dataset, while the results change little from 1e-3 to 1e-6. Besides, the performance is insensitive to the value change of margin parameter.

For the \textit{\textbf{parameter size}}, 
%it should be noted that our proposed network has the same size of 1,024 dimensions on each layer for different datasets, except the number of hidden units on the first layer in DBN, which is set according to the dimensional number of input features in different datasets. This design is for better generalization and reproducibility.
%Besides, 
we conduct experiments between our proposed approach and two best DNN-based methods namely CMDN and Corr-AE with the comparable sizes of parameter space on Wikipedia dataset taking CNN feature as input. 
Specifically, our CCL approach originally has about 21K parameters, while the CMDN and Corr-AE have about 17K and 10K parameters respectively. So we decrease the size of parameter space of CCL and CMDN, by reducing the number of layers and the number of hidden units on each layer to achieve the comparable parameter size with Corr-AE with about 10K parameters, which is called CCL-small and CMDN-small in Table \ref{table:para}.
%Specifically, we decrease the size of parameter space for the proposed CCL approach and our previous conference paper CMDN, which originally have about 21K and 17K parameters, by reducing the number of layers and the number of hidden units on each layer to achieve the close parameter size with Corr-AE with about 10K parameters. 
The experimental results are shown in Table \ref{table:para}. We can see that CCL-small still outperforms CMDN-small and Corr-AE with comparable parameter sizes. It shows that although retrieval accuracy can be improved by increasing parameter size, the size of parameter space is not the key point for the improvement. The effective modeling of intra-modality and inter-modality correlation makes great contribution to the final performance.

%As for the combination of coarse-grained and fine-grained information, we adopt a joint RBM to combine the coarse-grained and fine-grained intermediate representation for each modality, instead of straight-forward fusion strategies such as early fusion or late fusion, which aims to effectively learn the complementarity between these two kinds of representations by modeling the joint distribution over them for better fusion performance. 

\begin{table}[htb]
	\caption{Experiments on the parameter size between representative DNN-based methods.}
	\begin{center}
		\scalebox{0.9}{
			\begin{tabular}{|c|c|c|c|c|} 
				\hline
				\multirow{2}{*}{Dataset}&
				\multirow{2}{*}{Method} & \multicolumn{3}{c|}{MAP with CNN feature} \\
				\cline{3-5}
				& & Image$\rightarrow$Text & Text$\rightarrow$Image & Average  \\
				\hline
				
				\multirow{4}{*}{\begin{tabular}{c} Wikipedia \\ dataset \end{tabular}}
				& \textbf{CCL} & \textbf{0.504} & \textbf{0.457} & \textbf{0.481} \\
				
				&   \multirow{1}{*}{\begin{tabular}{c}\textbf{CCL-small}\end{tabular}} & \multirow{1}{*}{\textbf{0.490}} & \multirow{1}{*}{\textbf{0.444}}& \multirow{1}{*}{\textbf{0.467}} \\
%				& \multirow{1}{*}{\begin{tabular}{c}CMDN\end{tabular}} & \multirow{1}{*}{0.488} & \multirow{1}{*}{0.427}& \multirow{1}{*}{0.458} \\
				& \multirow{1}{*}{\begin{tabular}{c}CMDN-small\end{tabular}} & \multirow{1}{*}{0.463} & \multirow{1}{*}{0.408}& \multirow{1}{*}{0.436} \\
				& \multirow{1}{*}{\begin{tabular}{c}Corr-AE\end{tabular}} & \multirow{1}{*}{0.402} & \multirow{1}{*}{0.395}& \multirow{1}{*}{0.399} \\
				\hline

			\end{tabular} 
		}
	\end{center}
	\vspace{-0.2cm}
	\label{table:para}
\end{table}

\begin{table}[htb]
	\caption{Training and testing time for our CCL approach and compared methods.}
	\begin{center}
		\scalebox{0.85}{
			\begin{tabular}{|c|c|c|c|c|} 
				\hline
				\multirow{2}{*}{Method} & \multicolumn{2}{c|}{Training} & \multicolumn{2}{c|}{Testing} \\
				\cline{2-5}
				 & Wikipedia & NUS-WIDE & Wikipedia & NUS-WIDE\\
				\hline
				\textbf{Our CCL Approach} & \textbf{2794s} & \textbf{2958s} & \textbf{0.40s} & \textbf{281s}\\
				CMDN \cite{DBLP:conf/ijcai/PengHQ16} & 2102s & 2434s & 0.38s & 275s \\
				
				DCCA \cite{DBLP:conf/cvpr/YanM15} & 683s & 1636s & 0.42s & 289s\\
				Corr-AE \cite{feng12014cross} & 1051s & 1237s& 0.37s & 283s\\
				Multimodal DBN \cite{srivastava2012learning} & 1343s & 1523s& 0.39s & 275s\\
				Bimodal AE \cite{ngiam32011multimodal} & 974s & 1304s& 0.35s & 279s\\	
				JRL \cite{ZhaiTCSVT2014JRL} & 211s & 1426s& 0.26s & 75s\\
				GMM+HGLMM \cite{DBLP:conf/cvpr/KleinLSW15} & 235s & 803s& 0.30s & 143s\\
				Multi-label CCA \cite{DBLP:conf/iccv/RanjanRJ15} & 56s & 3924s& 0.29s & 110s\\
				MACC \cite{DBLP:conf/cvpr/TranBC16} & 62s & 2340s& 0.28s & 283s\\
				KCCA(Gaussian) \cite{DBLP:journals/neco/HardoonSS04} & 35s & 2018s& 0.35s & 2183s\\
				KCCA(Poly) \cite{DBLP:journals/neco/HardoonSS04} & 27s & 2228s& 0.31s & 2179s\\
				LGCFL \cite{DBLP:journals/tmm/KangXLXP15} & 8.2s & 28s & 0.26s & 65s\\
				CFA \cite{LiMM03CFA} & 7.5s & 8.7s& 0.76s & 264s\\
				CCA \cite{HotelingBiometrika36RelationBetweenTwoVariates} & 0.3s & 1.4s& 0.28s & 108s\\
				\hline
				
			\end{tabular} 
		}
	\end{center}
	\vspace{-0.2cm}
	\label{table:time}
\end{table}

\begin{table*}[htb]
	\caption{Baseline experiments on \textbf{Bi-modal Retrieval}, where \textbf{CCL (only coarse-grained)} means CCL with only the original instances as input, while \textbf{CCL (only fine-grained)} means CCL with only the patches of original instance as input. And \textbf{CCL (only intra-modality)} means CCL with only intra-modality separate representation learning, while \textbf{CCL (only inter-modality)} means CCL with only inter-modality separate representation learning. }
	\begin{center}
		\scalebox{0.9}{
			\begin{tabular}{|c|c|c|c|c|c|c|c|} 
				\hline
				\multirow{2}{*}{Dataset}&
				\multirow{2}{*}{Method} & \multicolumn{3}{c|}{MAP with hand-crafted feature} & \multicolumn{3}{c|}{MAP with CNN feature}\\
				\cline{3-8}
				& & Image$\rightarrow$Text & Text$\rightarrow$Image & Average & Image$\rightarrow$Text & Text$\rightarrow$Image & Average  \\
				\hline
				
				\multirow{5}{*}{\begin{tabular}{c} Wikipedia \\ dataset \end{tabular}}
				& \textbf{Our CCL Approach} & \textbf{0.418} & \textbf{0.359} & \textbf{0.389} & \textbf{0.504} & \textbf{0.457} & \textbf{0.481} \\
				
				&   \multirow{1}{*}{\begin{tabular}{c}CCL (only coarse-grained)\end{tabular}} & \multirow{1}{*}{0.399} & \multirow{1}{*}{0.329}& \multirow{1}{*}{0.364} & \multirow{1}{*}{0.457} & \multirow{1}{*}{0.429}& \multirow{1}{*}{0.443}\\
				& \multirow{1}{*}{\begin{tabular}{c}CCL (only fine-grained)\end{tabular}} & \multirow{1}{*}{0.393} & \multirow{1}{*}{0.328}& \multirow{1}{*}{0.361} & \multirow{1}{*}{0.456} & \multirow{1}{*}{0.400}& \multirow{1}{*}{0.428}\\
				%\hline
				
				&   \multirow{1}{*}{\begin{tabular}{c}CCL (only intra-modality)\end{tabular}} & \multirow{1}{*}{0.384} & \multirow{1}{*}{0.332}& \multirow{1}{*}{0.358} & \multirow{1}{*}{0.483} & \multirow{1}{*}{0.444}& \multirow{1}{*}{0.464}\\
				& \multirow{1}{*}{\begin{tabular}{c}CCL (only inter-modality)\end{tabular}} & \multirow{1}{*}{0.380} & \multirow{1}{*}{0.321}& \multirow{1}{*}{0.351} & \multirow{1}{*}{0.471} & \multirow{1}{*}{0.395}& \multirow{1}{*}{0.433}\\
				\hline
				
				\multirow{5}{*}{\begin{tabular}{c} NUS-WIDE-10K \\ dataset \end{tabular}} 
				&  \textbf{Our CCL Approach} & \textbf{0.400} & \textbf{0.401} & \textbf{0.401} & \textbf{0.506} & \textbf{0.535} & \textbf{0.521} \\
				
				&   \multirow{1}{*}{\begin{tabular}{c}CCL (only coarse-grained)\end{tabular}} & \multirow{1}{*}{0.371} & \multirow{1}{*}{0.356}& \multirow{1}{*}{0.364} & \multirow{1}{*}{0.483} & \multirow{1}{*}{0.516}& \multirow{1}{*}{0.500}\\
				& \multirow{1}{*}{\begin{tabular}{c}CCL (only fine-grained)\end{tabular}} & \multirow{1}{*}{0.368} & \multirow{1}{*}{0.358}& \multirow{1}{*}{0.363} & \multirow{1}{*}{0.414} & \multirow{1}{*}{0.436}& \multirow{1}{*}{0.425}\\
				
				&   \multirow{1}{*}{\begin{tabular}{c}CCL (only intra-modality)\end{tabular}} & \multirow{1}{*}{0.377} & \multirow{1}{*}{0.374}& \multirow{1}{*}{0.376} & \multirow{1}{*}{0.488} & \multirow{1}{*}{0.520}& \multirow{1}{*}{0.504}\\
				& \multirow{1}{*}{\begin{tabular}{c}CCL (only inter-modality)\end{tabular}} & \multirow{1}{*}{0.347} & \multirow{1}{*}{0.346}& \multirow{1}{*}{0.347} & \multirow{1}{*}{0.442} & \multirow{1}{*}{0.470}& \multirow{1}{*}{0.456}\\
				\hline
				
				\multirow{5}{*}{\begin{tabular}{c} Pascal Sentence \\ dataset \end{tabular}} 
				&  \textbf{Our CCL Approach} & \textbf{0.359} & \textbf{0.346} & \textbf{0.353} & \textbf{0.566} & \textbf{0.560} & \textbf{0.563} \\
				
				&   \multirow{1}{*}{\begin{tabular}{c}CCL (only coarse-grained)\end{tabular}} & \multirow{1}{*}{0.283} & \multirow{1}{*}{0.286}& \multirow{1}{*}{0.285} & \multirow{1}{*}{0.514} & \multirow{1}{*}{0.508}& \multirow{1}{*}{0.511}\\
				& \multirow{1}{*}{\begin{tabular}{c}CCL (only fine-grained)\end{tabular}} & \multirow{1}{*}{0.255} & \multirow{1}{*}{0.239}& \multirow{1}{*}{0.247} & \multirow{1}{*}{0.386} & \multirow{1}{*}{0.393}& \multirow{1}{*}{0.390}\\
				
				&   \multirow{1}{*}{\begin{tabular}{c}CCL (only intra-modality)\end{tabular}} & \multirow{1}{*}{0.296} & \multirow{1}{*}{0.272}& \multirow{1}{*}{0.284} & \multirow{1}{*}{0.521} & \multirow{1}{*}{0.520}& \multirow{1}{*}{0.521}\\
				& \multirow{1}{*}{\begin{tabular}{c}CCL (only inter-modality)\end{tabular}} & \multirow{1}{*}{0.217} & \multirow{1}{*}{0.217}& \multirow{1}{*}{0.217} & \multirow{1}{*}{0.420} & \multirow{1}{*}{0.383}& \multirow{1}{*}{0.402}\\
				\hline
				
				\multirow{5}{*}{\begin{tabular}{c} NUS-WIDE \\ dataset \end{tabular}} 
				&  \textbf{Our CCL Approach} & \textbf{0.513} & \textbf{0.489} & \textbf{0.501} & \textbf{0.671} & \textbf{0.676} & \textbf{0.674} \\
				
				&   \multirow{1}{*}{\begin{tabular}{c}CCL (only coarse-grained)\end{tabular}} & \multirow{1}{*}{0.502} & \multirow{1}{*}{0.466}& \multirow{1}{*}{0.484} & \multirow{1}{*}{0.665} & \multirow{1}{*}{0.670}& \multirow{1}{*}{0.668}\\
				& \multirow{1}{*}{\begin{tabular}{c}CCL (only fine-grained)\end{tabular}} & \multirow{1}{*}{0.485} & \multirow{1}{*}{0.451}& \multirow{1}{*}{0.468} & \multirow{1}{*}{0.595} & \multirow{1}{*}{0.604}& \multirow{1}{*}{0.600}\\
				
				&   \multirow{1}{*}{\begin{tabular}{c}CCL (only intra-modality)\end{tabular}} & \multirow{1}{*}{0.487} & \multirow{1}{*}{0.460}& \multirow{1}{*}{0.474} & \multirow{1}{*}{0.652} & \multirow{1}{*}{0.660}& \multirow{1}{*}{0.656}\\
				& \multirow{1}{*}{\begin{tabular}{c}CCL (only inter-modality)\end{tabular}} & \multirow{1}{*}{0.427} & \multirow{1}{*}{0.400}& \multirow{1}{*}{0.414} & \multirow{1}{*}{0.515} & \multirow{1}{*}{0.519}& \multirow{1}{*}{0.517}\\
				\hline
				
			\end{tabular} 
		}
	\end{center}
	\vspace{-0.5cm}
	\label{table:Baseline}
\end{table*}

\begin{table}[htb]
	\caption{Baseline experiments of \textbf{Image Annotation} on Flickr-30K and MS-COCO datasets.}%, where \textbf{CCL (only coarse-grained)} means CCL with only the original instances as input, while \textbf{CCL (only fine-grained)} means CCL with only the patches of original instance as input. And \textbf{CCL (only intra-modality)} means CCL with only intra-modality separate representation learning, while \textbf{CCL (only inter-modality)} means CCL with only inter-modality separate representation learning. }
	\begin{center}
		\scalebox{0.69}{
			\begin{tabular}{|c|c|c|c|c|c|c|c|} 
				\hline
				\multirow{2}{*}{Dataset}&
				\multirow{2}{*}{Method} & \multicolumn{3}{c|}{Hand-crafted feature} & \multicolumn{3}{c|}{CNN feature}\\
				\cline{3-8}
				& & R@1 & R@5 & R@10 & R@1 & R@5 & R@10  \\
				\hline
				
				\multirow{5}{*}{\begin{tabular}{c} Flickr-30K \\ dataset \end{tabular}}
				&\textbf{Our CCL Approach} & \textbf{0.088} & \textbf{0.276} & \textbf{0.393} & \textbf{0.377} & \textbf{0.694} & \textbf{0.811} \\
				
				&   \multirow{1}{*}{\begin{tabular}{c}CCL (only coarse-grained)\end{tabular}} & \multirow{1}{*}{0.059} & \multirow{1}{*}{0.192}& \multirow{1}{*}{0.294} & \multirow{1}{*}{0.286} & \multirow{1}{*}{0.587}& \multirow{1}{*}{0.719}\\
				& \multirow{1}{*}{\begin{tabular}{c}CCL (only fine-grained)\end{tabular}} & \multirow{1}{*}{0.021} & \multirow{1}{*}{0.100}& \multirow{1}{*}{0.171} & \multirow{1}{*}{0.170} & \multirow{1}{*}{0.436}& \multirow{1}{*}{0.581}\\
				
				&   \multirow{1}{*}{\begin{tabular}{c}CCL (only intra-modality)\end{tabular}} & \multirow{1}{*}{0.040} & \multirow{1}{*}{0.133}& \multirow{1}{*}{0.220} & \multirow{1}{*}{0.312} & \multirow{1}{*}{0.641}& \multirow{1}{*}{0.753}\\
				& \multirow{1}{*}{\begin{tabular}{c}CCL (only inter-modality)\end{tabular}} & \multirow{1}{*}{0.029} & \multirow{1}{*}{0.125}& \multirow{1}{*}{0.212} & \multirow{1}{*}{0.232} & \multirow{1}{*}{0.544}& \multirow{1}{*}{0.694}\\
				\hline
				
				\multirow{5}{*}{\begin{tabular}{c} MS-COCO \\ dataset \end{tabular}} 
				&  \textbf{Our CCL Approach} & \textbf{0.063} & \textbf{0.196} & \textbf{0.298} & \textbf{0.186} & \textbf{0.474} & \textbf{0.625} \\
				
				&   \multirow{1}{*}{\begin{tabular}{c}CCL (only coarse-grained)\end{tabular}} & \multirow{1}{*}{0.048} & \multirow{1}{*}{0.154}& \multirow{1}{*}{0.241} & \multirow{1}{*}{0.127} & \multirow{1}{*}{0.364}& \multirow{1}{*}{0.510}\\
				& \multirow{1}{*}{\begin{tabular}{c}CCL (only fine-grained)\end{tabular}} & \multirow{1}{*}{0.030} & \multirow{1}{*}{0.100}& \multirow{1}{*}{0.162} & \multirow{1}{*}{0.102} & \multirow{1}{*}{0.304}& \multirow{1}{*}{0.436}\\
				
				&   \multirow{1}{*}{\begin{tabular}{c}CCL (only intra-modality)\end{tabular}} & \multirow{1}{*}{0.058} & \multirow{1}{*}{0.177}& \multirow{1}{*}{0.264} & \multirow{1}{*}{0.147} & \multirow{1}{*}{0.406}& \multirow{1}{*}{0.556}\\
				& \multirow{1}{*}{\begin{tabular}{c}CCL (only inter-modality)\end{tabular}} & \multirow{1}{*}{0.026} & \multirow{1}{*}{0.096}& \multirow{1}{*}{0.160} & \multirow{1}{*}{0.121} & \multirow{1}{*}{0.340}& \multirow{1}{*}{0.485}\\
				\hline

			\end{tabular} 
		}
	\end{center}
	\vspace{-0.5cm}
	\label{table:Baseline3}
\end{table}

\begin{table}[htb]
	\caption{Baseline experiments of \textbf{Image Retrieval} on Flickr-30K and MS-COCO datasets.}%, where \textbf{CCL (only coarse-grained)} means CCL with only the original instances as input, while \textbf{CCL (only fine-grained)} means CCL with only the patches of original instance as input. And \textbf{CCL (only intra-modality)} means CCL with only intra-modality separate representation learning, while \textbf{CCL (only inter-modality)} means CCL with only inter-modality separate representation learning.}
	\begin{center}
		\scalebox{0.69}{
			\begin{tabular}{|c|c|c|c|c|c|c|c|} 
				\hline
				\multirow{2}{*}{Dataset}&
				\multirow{2}{*}{Method} & \multicolumn{3}{c|}{Hand-crafted feature} & \multicolumn{3}{c|}{CNN feature}\\
				\cline{3-8}
				& & R@1 & R@5 & R@10 & R@1 & R@5 & R@10  \\
				\hline
				
				\multirow{5}{*}{\begin{tabular}{c} Flickr-30K \\ dataset \end{tabular}}
				& \textbf{Our CCL Approach} & \textbf{0.090} & \textbf{0.253} & \textbf{0.361} & \textbf{0.373} & \textbf{0.684} & \textbf{0.800} \\

				&   \multirow{1}{*}{\begin{tabular}{c}CCL (only coarse-grained)\end{tabular}} & \multirow{1}{*}{0.054} & \multirow{1}{*}{0.179}& \multirow{1}{*}{0.272} & \multirow{1}{*}{0.255} & \multirow{1}{*}{0.567}& \multirow{1}{*}{0.692}\\
				& \multirow{1}{*}{\begin{tabular}{c}CCL (only fine-grained)\end{tabular}} & \multirow{1}{*}{0.021} & \multirow{1}{*}{0.098}& \multirow{1}{*}{0.172} & \multirow{1}{*}{0.172} & \multirow{1}{*}{0.422}& \multirow{1}{*}{0.569}\\
				
				&   \multirow{1}{*}{\begin{tabular}{c}CCL (only intra-modality)\end{tabular}} & \multirow{1}{*}{0.033} & \multirow{1}{*}{0.135}& \multirow{1}{*}{0.205} & \multirow{1}{*}{0.310} & \multirow{1}{*}{0.613}& \multirow{1}{*}{0.745}\\
				& \multirow{1}{*}{\begin{tabular}{c}CCL (only inter-modality)\end{tabular}} & \multirow{1}{*}{0.028} & \multirow{1}{*}{0.116}& \multirow{1}{*}{0.205} & \multirow{1}{*}{0.228} & \multirow{1}{*}{0.544}& \multirow{1}{*}{0.688}\\
				\hline
				
				\multirow{5}{*}{\begin{tabular}{c} MS-COCO \\ dataset \end{tabular}} 
				&  \textbf{Our CCL Approach} & \textbf{0.064} & \textbf{0.197} & \textbf{0.296} & \textbf{0.196} & \textbf{0.469} & \textbf{0.623} \\
				
				&   \multirow{1}{*}{\begin{tabular}{c}CCL (only coarse-grained)\end{tabular}} & \multirow{1}{*}{0.047} & \multirow{1}{*}{0.162}& \multirow{1}{*}{0.247} & \multirow{1}{*}{0.132} & \multirow{1}{*}{0.358}& \multirow{1}{*}{0.505}\\
				& \multirow{1}{*}{\begin{tabular}{c}CCL (only fine-grained)\end{tabular}} & \multirow{1}{*}{0.027} & \multirow{1}{*}{0.100}& \multirow{1}{*}{0.160} & \multirow{1}{*}{0.107} & \multirow{1}{*}{0.299}& \multirow{1}{*}{0.430}\\
				
				&   \multirow{1}{*}{\begin{tabular}{c}CCL (only intra-modality)\end{tabular}} & \multirow{1}{*}{0.056} & \multirow{1}{*}{0.181}& \multirow{1}{*}{0.273} & \multirow{1}{*}{0.150} & \multirow{1}{*}{0.402}& \multirow{1}{*}{0.547}\\
				& \multirow{1}{*}{\begin{tabular}{c}CCL (only inter-modality)\end{tabular}} & \multirow{1}{*}{0.026} & \multirow{1}{*}{0.104}& \multirow{1}{*}{0.171} & \multirow{1}{*}{0.119} & \multirow{1}{*}{0.330}& \multirow{1}{*}{0.473}\\
				\hline

			\end{tabular} 
		}
	\end{center}
	\vspace{-0.5cm}
	\label{table:Baseline4}
\end{table}

\subsubsection{Execution Time}

We measure the execution time for our proposed CCL approach as well as all the compared methods. 
It should be noted that we measure the training and testing time for different methods without the time for data preprocessing such as feature extraction and patch segmentation. 
Besides, we adopt small-scale Wikipedia dataset and large-scale NUS-WIDE dataset, and use CNN feature as input. 
The execution time is recorded on a PC with i7-5930K CPU @ 3.50GHz, 64GB memory and one GPU of NVIDIA Titan X with 12GB memory. The \textit{\textbf{training time}} is shown in Table \ref{table:time}. 
We can see that the training time of all methods increases with the number of training instances increasing, especially for some traditional methods such as JRL and KCCA due to the large amount of calculation on matrix multiplication. % as well as the high time complexity with $O(n^2)$ at least. 
Besides, among the DNN-based methods, the training time of Bimodal AE, Multimodal DBN and Corr-AE is close to each other and less than our proposed CCL approach and CMDN. It is because our CCL approach and CMDN have more components to train for different purposes, such as different subnetworks for coarse-grained and fine-grained representation learning and the multi-grained fusion subnetwork in CCL approach, while the compared DNN-based methods ignore the complementary fine-grained clues provided by their patches. 
On \textit{\textbf{testing time}}, all the methods including our proposed CCL approach have almost the same testing time on small-scale Wikipedia dataset. While on large-scale NUS-WIDE dataset, the time has some difference mainly because the dimensional numbers of common representation generated from different methods vary greatly, which leads to different costs in distance measurement with the increasing number of testing instances. For example, KCCA has the longest testing time due to its high-dimensional common representation with about 10,000 dimensions, which is much higher than other compared methods having less than 1,024 dimensions. 
From the above results, we can see that the training stage is the key point for our proposed CCL approach to achieve the best accuracy, while it costs little extra time in the testing stage.

\subsubsection{Baseline Experiments}
Tables \ref{table:Baseline} to \ref{table:Baseline4} show the accuracies of our CCL approach and the baseline approaches. CCL (only coarse-grained) means only the coarse-grained original instances are used to generate common representation, while CCL (only fine-grained) means only the fine-grained patches are adopted. Due to only the coarse-grained or fine-grained information is adopted, we drop the joint RBM which is used to fuse multi-grained representation, and the other parts of the network keep exactly the same in our CCL approach for fair comparison. The coarse-grained and fine-grained information can be effectively integrated by CCL approach to make cross-modal correlation more precise, which leads to better accuracy than CCL (only coarse-grained) and CCL (only fine-grained) that use single-grained information merely.

%CCL-global has higher accuracy than CCL-patch, which indicates that the coarse-grained information provided by original instance is more useful than the fine-grained information for cross-modal correlation modeling. While our CCL approach which integrates the coarse-grained instances and fine-grained patches, achieves higher accuracy than the CCL-global and CCL-patch that only use single type of information, which shows that the patches still contain complementary fine-grained clues to the original instances to make cross-modal correlation more precise.

Besides, we also show the baseline experiments to verify the effectiveness of combination of intra-modality and inter-modality correlation in the first learning stage. CCL (only intra-modality) means CCL with only intra-modality reconstruction learning error $L_r$ in Eq.(\ref{equ:ls}), while CCL (only inter-modality) means CCL with only inter-modality correlation learning error $L_c$ in Eq.(\ref{equ:ls}) for separate representation learning. 
%We can see that the results of CCL-intra outperforms CCL-inter, which means that the intra-modality information plays a more important role in the first learning stage, 
We can see that learning intra-modality and inter-modality correlation simultaneously by joint optimization achieves higher MAP score than learning only one of them, which indicates the complementarity of the two kinds of information, and it can be effectively exploited by our CCL approach to achieve better accuracy.
%Similar to the previous baseline experiment, only CCL-intra or CCL-inter can outperform most of the compared methods.

The above baseline experimental results verify the contribution of each component in our CCL approach. First, inter-modality correlation can provide rich complementary context to intra-modality correlation for learning better separate representation, which can capture the important hints to boost common representation learning. Second, fusion of complementary coarse-grained instances and fine-grained patches can lead to more accurate cross-modal common representation.

\section{Conclusion}

In this paper, a cross-modal correlation learning approach has been proposed with multi-grained fusion by hierarchical network. 
In the first learning stage, separate representation is learned by jointly optimizing intra-modality and inter-modality correlation, which can effectively capture the intrinsic correlation and rich complementary information in different modalities. 
In the second learning stage, a multi-task learning strategy, which adaptively balances intra-modality semantic category constraints and inter-modality pairwise similarity constraints, is adopted to fully exploit the intrinsic relevance between them. 
Besides, the coarse-grained instances and fine-grained patches are integrated to make cross-modal correlation more precise. 
Comprehensive experimental results show effectiveness of our CCL approach compared with 13 state-of-the-art methods on 6 widely-used cross-modal datasets. 

The future work lies in two aspects: First, we will focus on learning better fine-grained representation with more effective and precise segmentation methods. Second, we will attempt to apply semi-supervised regularization into our framework to make full use of the unlabeled data. Both of them will be employed to further boost the cross-modal retrieval accuracy.

% or


%\appendices
%% you can choose not to have a title for an appendix
%% if you want by leaving the argument blank
%\section{}

%\section*{Acknowledgment}
%This work was supported by National Natural Science Foundation of China under Grants 61371128 and 61532005.

\ifCLASSOPTIONcaptionsoff
  \newpage
\fi



\bibliographystyle{IEEEtran}
\bibliography{ijcai16}
%i
\begin{IEEEbiography}[{\includegraphics[width=1in,height=1.25in,clip,keepaspectratio]{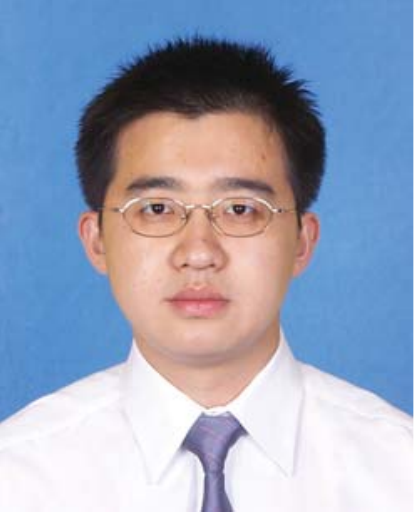}}]{Yuxin Peng}
	is the professor of Institute of Computer Science and Technology (ICST), Peking University, and the chief scientist of 863 Program (National Hi-Tech Research and Development Program of China). He received the Ph.D. degree in computer application technology from Peking University in Jul. 2003. After that, he worked as an assistant professor in ICST, Peking University. He was promoted to associate professor and professor in Peking University in Aug. 2005 and Aug. 2010 respectively. In 2006, he was authorized by the ``Program for New Star in Science and Technology of Beijing'' and the ``Program for New Century Excellent Talents in University (NCET)''. He has published over 100 papers in refereed international journals and conference proceedings, including IJCV, TIP, TCSVT, TMM, PR, ACM MM, ICCV, CVPR, IJCAI, AAAI, etc. He led his team to participate in TRECVID (TREC Video Retrieval Evaluation) many times. In TRECVID 2009, his team won four first places on 4 sub-tasks of the High-Level Feature Extraction (HLFE) task and Search task. In TRECVID 2012, his team gained four first places on all 4 sub-tasks of the Instance Search (INS) task and Known-Item Search (KIS) task. In TRECVID 2014, his team gained the first place in the Interactive Instance Search task. His team also gained both two first places in the INS task of TRECVID 2015 and 2016. Besides, he won the first prize of Beijing Science and Technology Award for Technological Invention in 2016 (ranking first). He has applied 34 patents, and obtained 16 of them. His current research interests mainly include cross-media analysis and reasoning, image and video analysis and retrieval, and computer vision.
\end{IEEEbiography}

\begin{IEEEbiography}[{\includegraphics[width=1in,height=1.25in,clip,keepaspectratio]{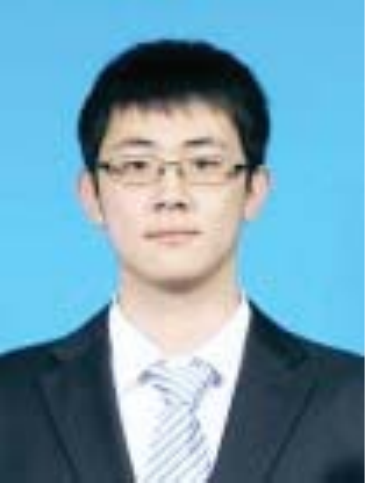}}]{Jinwei Qi}
	received the B.S. degree in computer science and technology from Peking University, in Jul. 2016. He is currently pursuing the M.S. degree with the Institute of Computer Science and Technology (ICST), Peking University. His current research interests include cross-media retrieval and deep learning.
\end{IEEEbiography}
\begin{IEEEbiography}[{\includegraphics[width=1in,height=1.25in,clip,keepaspectratio]{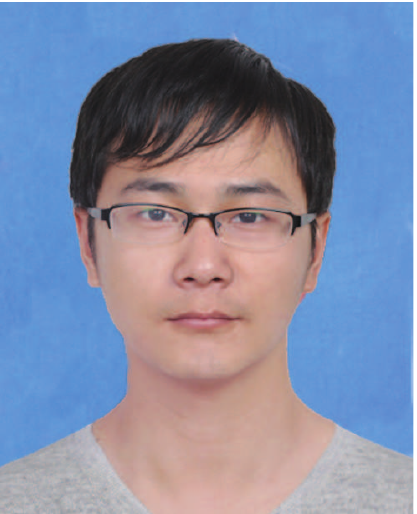}}]{Xin Huang}
	received the B.S. degree in computer science and technology from Peking University, in Jul. 2014. He is currently pursuing the Ph.D. degree in the Institute of Computer Science and Technology (ICST), Peking University. His research interests include cross-media reasoning and deep learning.
\end{IEEEbiography}
\begin{IEEEbiography}[{\includegraphics[width=1in,height=1.25in,clip,keepaspectratio]{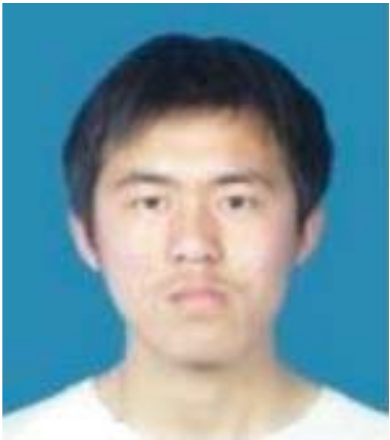}}]{Yuxin Yuan}
	received the B.S. degree in electronic science and engineering from Nanjing University, in Jul. 2015. He is currently pursuing the M.S. degree in the Institute of Computer Science and Technology (ICST), Peking University. His current research interests include cross-media retrieval and deep learning.
\end{IEEEbiography}
\end{document}